%% file: cs.tex
\documentclass[vecphys,natbib]{svmult}

\usepackage{ergebnisse} 

\usepackage{makeidx}         
\usepackage{graphicx}        
\usepackage{multicol}        
\usepackage[bottom]{footmisc}

\makeindex             

\begin{document}

\title{Galaxy Collisions - Dawn of a New Era}
\titlerunning{Galaxy Collisions} 
\author{Curtis Struck}
\institute{}
%
%
\maketitle

\begin{abstract}

The study of colliding galaxies has progressed rapidly in the last few
years, driven by observations with powerful new ground and space-based
instruments. These instruments have used for detailed studies of
specific nearby systems, statistical studies of large samples of
relatively nearby systems, and increasingly large samples of high
redshift systems. Following a brief summary of the historical context,
this review attempts to integrate these studies to address the
following key issues. What role do collisions play in galaxy
evolution, and how can recently discovered processes like downsizing
resolve some apparently contradictory results of high redshift
studies? What is the role of environment in galaxy collisions?  How is
star formation and nuclear activity orchestrated by the large scale
dynamics, before and during merger? Are novel modes of star formation
involved? What are we to make of the association of ultraluminous
X-ray sources with colliding galaxies? To what do degree do mergers
and feedback trigger long-term secular effects? How far can we push
the archaeology of individual systems to determine the nature of
precursor systems and the precise effect of the interaction? Tentative
answers to many of these questions have been suggested, and the
prospects for answering most of them in the next few decades are good.

\end{abstract}


\section{Introduction: Some Past Highlights and Current Issues}
\label{sec:1}

\subsection{Early Days}
\label{sec:2}

The study of galaxy collisions is not an ancient one; Erik Holmberg,
Fritz Zwicky, and a few others did quite a bit of work relevant to
colliding galaxies before the 1950s, but that decade opened and closed
with two landmark papers. Thus, we can justify taking it as the first
decade of general interest in the subject, and view the earlier work
as pioneering. The first of the two papers, by \citet{spi51}, revived
Zwicky's suggestion that collisions would be frequent within dense
galaxy clusters, and considered what would happen in direct collisions
between two galaxies. Specifically, they correctly argued that the
stellar distribution might be only moderately disturbed, while strong
shock waves could push the interstellar gas out of the galaxies. Their
primary conclusion was that this process could account for the
scarcity of late-type spiral galaxies with substantial ongoing star
formation in clusters. For the first time galaxy collisions were seen
to have an important role in galaxy evolution.

The second landmark paper was Zwicky's review of his extensive imagery
of morphologically peculiar galaxies, together with arguments that
many of these peculiarities were caused by tidal forces in collisions
\citep{zwi59}. At the time, Zwicky's theory seemed doubtful, since
collisions between the widely separated ``island universes'' were
deemed improbable. Moreover, his arguments were generally
semi-quantitative, and so, not compelling.

In the middle of this first decade \citet{baa54} suggested that one of
the most prominent members of the newly discovered class of radio
galaxies, Cygnus A, was in collision. Thus, we already have the first
hints of many of the most important themes in this field, including:
the generation of unique tidal morphologies, induced nuclear activity,
induced star formation, the important role of collisions in galaxy
evolution, and the dependence of these effects on the clustering
environment. In this review I will focus on the last few items -
induced star formation, galaxy evolution, and environmental effects -
and say relatively little about the first two and many other related
topics.

In the second decade, much of the work was of a more detailed, and
sometimes indirect nature, which I cannot review here (see e.g.,
\citealt{str99a}). The major exception to this generalization was
publication of Arp's pictoral atlas of more than 300 peculiar galaxies
\citep{arp66}. Arp derived the atlas objects from the Palomar sky
survey. In the following decades this atlas became the standard 'field
guide' for workers in this field. The Arp galaxies were arranged in
categories somewhat like Zwicky's, but with many more examples, and
excellent photographic images. One psychological effect of so many
images may well have been to make the peculiar galaxies seem less like
freakish rarities, and more like zoological families in need of
explanation.

\subsection{The 1970s}
\label{sec:3}

\citet{too72} took a giant step toward these
explanations. Their numerical models were not the first, and were
simple by modern standards, but they were more extensive than previous
efforts. They were able to account for many of the Arp atlas forms in
detail, thereby making a strong case for the collisional origins
theory. They also made a number of important predictions for
observation, such as that strong tidal waves would lead to enhanced
star formation and gas transfer to nuclear regions, which could fuel
nuclear activity. These would become dominant themes in subsequent
work.

\begin{figure}
\centering
%
\caption{Multiwaveband images of several well-known merger remnants
(courtesy D. B. Sanders and I. F. Mirabel): a) NGC 4038/39 (Arp 244,
``The Antennae''), b) NGC 7252 (Arp 226, ``Atoms for Peace''), c) IRAS
19254-7245 (``The Super-Antennae''), d) IC 4553/54 (Arp 220). An
optical image shown in greyscale, HI (21 cm) surface intensity shown
by contours, and K band (2.2 mm) shown in insets. Scale-bar represents
20 kpc in each case; see \citet{san96} for details.}
\label{fig:1}
\end{figure}

However, Toomre and Toomre's models did not directly account for the
'messiest' objects in the Arp atlas (see examples in Figure 1). Alar
Toomre returned to these objects in his contribution to the seminal
Yale galaxies conference \citep{too77}. He pointed out that the
earlier models had not included the effect of Chandrasekhar's
dynamical friction \citep{cha43}, and showed that the effect would
draw the colliding galaxies into a merger. He further considered how
merger remnants would evolve, and how they would appear
observationally. This lead him to some radical conclusions for that
time, that mergers between comparable large spiral galaxies could lead
to the formation of elliptical galaxies, and that reasonable
extrapolation of the statistics of such collisions suggested that a
large fraction of ellipticals could be formed this way. The debate
still continues on many aspects of this scenario, but it immediately
had an important effect. Toomre had opened the door to the possibility
that collisions were the dominant factor in the evolution of an
important class of galaxies. Collisions were more than just a means of
accounting for rare freaks, or a specialized process peculiar to the
environment of dense clusters.

Other important developments in the 1970s included the work of
\citet{lar78}, who suggested that Arp atlas galaxies had a wider range
of optical colors and star formation rates (SFRs) than the more normal
Hubble atlas \citep{san61} galaxies. Extension of that work suggested
that infrared colors would provide even more sensitive indications of
varying SFRs \citep{str78}. Ever increasing evidence that galaxies
(and groups and clusters) possessed massive dark halos (see
\citet{sof01} for a history of rotation curve studies) completely
changed our understanding of what a galaxy is. The ten-fold increase
of galaxy masses and sizes in the new picture provided an explanation
of why collisions could be common, despite the great separations of
the visible parts of galaxies. Their cross section were much larger
than previously thought, and collision partners were born bound
together in larger entities.

\subsection{The 1980s and Early 1990s}
\label{sec:4}

This period saw expansion of the field into many new directions, with
a number of major developments that defined the current epoch. One of
the highest points was the discovery of ultraluminous far-infrared
galaxies (ULIRGs) with the observations of the IRAS satellite (see
reviews of \citealt{soi87} and \citealt{san96}). This discovery set off
a gold rush of studies of these objects, as illustrated by the papers
of the 1986 Pasadena meeting \citep{lon86} and the 1989 Alabama
meeting \citep{sul89}, and which continues to some degree up to the
present. A primary focus of most ULIRG papers has been the relative
role of nuclear starbursts versus active nuclei in generating the huge
emissions. This is a difficult question to answer because both are
usually buried deeply in the gas and dust of the merger remnant; most
observational techniques give only indirect clues. While elucidating
the connection between starburst and nuclear activity is very
important, the ULIRGs and their somewhat less luminous cousins, the
LIRGs, offer a wealth of other information on questions of galaxy
evolution.

A second focus of ULIRG studies was the determination of what sort of
remnant would ultimately emerge from a major merger. ULIRGs could be
seen as the missing link in Toomre's theory of elliptical formation
from major mergers. They are recent mergers with prodigious amounts of
star formation, which might eventually either consume or heat and
disperse the gas, as required by the theory. The fact that the old
star surface brightness profile approximated the de Vaucouleurs
profile characteristic of ellipticals in the inner regions of some
ULIRGs, despite the presence of tidal distortions in the outer parts,
gave further support to the theory.

This was generally a period of rapid development of numerical
models. It began with the publication of the first fully
self-consistent three-dimensional models of galaxy collisions followed
through the merger (see review of \citealt{bar92}). In these models
the galaxies were of comparable size and consisted of a single
spheroidal component, i.e. like two elliptical galaxies without dark
haloes. They showed that mergers occurred much more quickly than
expected, as orbital energy was efficiently channeled into internal
collective modes. They also revealed the rapid appearance of a de
Vaucouleurs surface density profile in some major merger
remnants. This profile can be viewed as a kind of meta-stable state,
resulting from the prompt relaxation of collective modes. Its
appearance in ULIRGs indicated agreement between observations and
models, and provided more support for the ellipticals from mergers
theory.

By the end of this period the state of the numerical art had advanced
to self-consistent merger models of galaxies with stellar disk, gas
disk, and dark halo components \citep{bar92}. These models showed that
different galaxy components behaved somewhat differently during the
(major) merger process, with dynamically hot halo components generally
merging more quickly than the disk components. Even more exciting from
the point of view of ULIRG studies, the models showed that a fraction
of the gas carried much of the angular momentum out into extended
tidal structures, while the rest of the gas fell into a small volume
in the remnant center. This mass of highly compressed gas could
readily fuel ULIRG superstarbursts.

This period did not see many models of mini or micro mergers, in part
because ULIRGs and major mergers were the focus, but also because
adequate numerical resolution of small companions was
difficult. Another lacuna of modeling in this period was realistic gas
dynamics; most models used either an isothermal equation of state for
the gas or 'sticky particle' algorithms with phenomenological
collision rules between particles representing gas clouds. Cooling,
heating and stellar feedback processes were not generally included,
(but see e.g., \citealt{app87b}).

Alongside the major thrusts of merger studies several quiet
revolutions occurred in this period. One of these was based on the
sensitive mapping of atomic hydrogen in galaxies generally, as well as
collisional systems, by many observers using the Westerbork array, and
later the VLA (Very Large Array of the National Radio Astronomy
Observatory). These observations first made clear that the gas disks
of typical disk galaxies were much larger than the stellar, and then
as one might have expected, that these extended gas disks were more
strongly affected by collisional encounters than the inner stellar
disks. It soon became clear that such observations were essential for
determing the full extent of tidal tails and bridges. HI mapping also
provides a map of the line-of-sight velocities of the gas. Kinematic
maps provide us a view in a third dimension of the six dimensional
position-velocity space, and this information is usually crucial to
the success of models of individual systems, thereby to detailed tests
of collision theory. The accomplishments of the VLA were summarized at
a recent symposium \citep{hib01}, and a valuable legacy of that
meeting was the creation of the HI Rogue's Gallery website of
colliding galaxy HI maps by J. Hibbard
(www.nrao.edu/astrores/Hirogues/).

Another discovery that can be described as revolutionary is that tidal
interactions can induce the formation of a bar component out of disk
material. This was shown by the numerical models of \citet{nog87}, and
studied in detail by Athanassoula (see review of \citet{ath04} and
references therein). This result is important because bars transfer
angular momentum outward in the disk, and so can drive gas into the
central regions before merger. The bar can also drive spiral density
waves. Both the increased central gas concentration and the bar/spiral
waves can induce star formation.

We will examine the question of SF induced before merging in more
detail below. However, we should note here that Keel, Kennicutt and
collaborators carried out an extensive program of H$\alpha$ imagery
and spectra of both collisional systems and of a control sample
(\citealt{kee85}, \citealt{ken87}). They found indications of enhanced
SF in the collision sample, and particularly of SF enhancements in
galaxy cores which were kinematically disturbed. On larger scales,
\citet{scho90} observed the broad band colors of a sample of tidal
bridges, plumes and tails, and found that while SF in these structures
was not especially strong, it did continue after their formation. This
is somewhat surprising given the great extent of many of these
structures, which would seem to imply diminished gas densities and SF.

In his continuing studies of putative merger remnant-to-elliptical
systems, Schweizer also discovered large, young star clusters or dwarf
galaxies formed in tidal tails, most notably in the ``Antennae''
system \citep{schw83}. These discoveries would inspire a great deal of
new work in the 1990s and the present decade. More generally,
Schweizer's detailed, multi-waveband studies of specific merger
remnants, whose appearance suggested that they were on the road to
becoming ellipticals, advanced Toomre's merger theory (see
\citet{schw98} and references therein).

As a final example of quiet revolutions of the 1980s I would include
the extensions to dynamical friction theory by \citet{tre84}, and the
application of the new theory to the evolution of galactic bars
\citep{wei85}. The classical \citet{cha43} theory was too idealized to
account for the frictional effects in major mergers, and even more so
in the case of a ``sinking satellite'' orbiting outside of, but
interacting with the disk of the primary galaxy. The Tremaine and
Weinberg theory included the collective effects not accounted for in
the classical theory, and is able to account for the rapidity of major
mergers seen in numerical models.

Even beyond these revolutionary examples the tapestry of colliding
galaxy studies also grew with the addition of more new threads in this
period. These included studies of many specific types of collisional
system, such as: colliding ring galaxies (see review of
\citealt{app87a}), polar rings (see review of \citealt{spa02}), ocular
ovals \citep{elm91}, and shell galaxies (e.g.,
\citealt{her88}). Numerical modeling demonstrated how these
distinctive morphologies could be produced in collisions, and thus
confirmed earlier conjectures on the broad scope of collision
theory. In addition, distinctive morphologies were generally found to
be the result of a relatively narrow set of collision
parameters. Examples in each class can be viewed as a set of related
natural experiments, seen at different times and with slightly
different initial parameter values, which have the potential to
provide much insight into difficult or obscure collision processes
(e.g., hydrodynamic or SF processes).

\subsection{Key Issues Up to the Present}
\label{sec:5}

The 1990s saw continued rapid expansion of the field, driven in part
by new ground and satellite-based instrumentation, and by rapidly
increasing computer power. It is very difficult to summarize the
accomplishments of that decade briefly. Queries to NASA's
Astrophysical Data System show that the number of literature papers
with abstracts containing the words ``galaxy'' and ``collision'' grew
very rapidly with each decade: 27 (1950s), 75 (1960s), 326 (1970s),
826 (1980s), 1413 (1990s). Similar increases in the number of studies
in the related fields of galaxy formation and galaxy evolution at high
redshift make the task even more difficult. In this review we will
focus our attention on key issues relating to star formation and
galaxy evolution.

It is clear that over the second half of the 20th century this field
has gone from bare beginnings to a considerable maturity, providing
answers to some of its most important questions and early
paradoxes. Yet many questions remain, including some that have been
common threads through the whole history of the subject, and which are
connected to the deepest questions in astrophysics. For reference in
the rest of this review, I list here some of the most important ones.

{\bf{1.}} How do collisions and interactions affect galaxy evolution
overall?  More precisely, what are the relative roles of major and
minor mergers in building galaxies? This question is related to that
of how galaxies form, since major mergers are very important in
hierarchical build-up models, and negligible in monolithic collapse
models of galaxy formation.

{\bf{2.}} How does the answer to the previous question depend on
environment?  How do collisions differ in cluster, group, or nearly
isolated environments? Some partial answers to these questions have
been known for a long time. For example, collisions between field
galaxies are very different from those between cluster galaxies
because the latter have typical relative velocities of thousands of
km/s versus velocities of hundreds of km/s in the former case. High
velocity collisions can remove interstellar gas and produce moderate
tidal distortions, but are unlikely to result in merger, while mergers
are generally inevitable in the lower velocity collisions in
groups. Research over the last few decades has provided a great deal
of information on these questions, and it has become clear that
environment plays a very large role in determining the nature of
collisions that can occur, and the relative importance of galaxy
collisions versus other evolutionary processes (like gas sweeping in
dense clusters).

{\bf{3.}} How do the large-scale dynamics of collisions and
interactions orchestrate star formation (SF) and nuclear activity,
which are inherently small scale processes? The clear answer from the
1980s is that activity is induced by dumping a great deal of gas into
the central regions of major merger remnants. Major mergers may be the
way to make most of the stars in a significant fraction of early type
galaxies, but they are a rare event in the world of galaxy collisions,
and the question remains for other types of collision. Related
questions include: when do galactic winds and fountains result from
interaction induced SF, and what feedback role do they play in the
subsequent SF?

{\bf{4.}} To what degree do mergers trigger long-term (more than 1.0
Gyr) secular processes? Examples include the long-term effects of
collisionally induced bar components, and the fallback of large scale
tidal structures.

{\bf{5.}} How far can we push the archaeology of individual systems?
Do enough clues remain to determine the morphology of the precursor
galaxies, and decipher the details of the interaction up to the
present?

In the remainder of this review we will consider how developments in
the last decade and the near future help to answer these
questions. The first three sets of questions include the key questions
of this review. The last two push beyond its scope, and I will not
treat them in any detail, despite their intrinsic interest.

\section{Induced Star Formation and Winds}
\label{sec:6}
\subsection{Star Formation Processes in Interactions}
\label{sec:7}

Star formation induced by galaxy collisions appears similar to SF in
isolated galaxies in several ways. Before merging it is often
concentrated in spiral waves or bars, and tidal tails often look like
extensions of the spirals. Both before and after merger it is often
concentrated in nuclear starbursts. These can be orders of magnitude
stronger than core bursts in isolated galaxies, but they can appear
qualitatively similar.

However, there are theoretical reasons to think that the nature of
collisionally induced SF is very different from that in isolated
disks. In isolated star-forming disks there is evidence that SF, and
gas disk structure, are self-regulated by energy and momentum
feedbacks from young SF regions (e.g., \citealt{ken89}). The
self-regulation processes work to maintain a gas surface density close
to the threshold for local gravitational instability throughout the
disk. SF is usually concentrated in grand design or flocculent phase
waves, which compress the gas, pushing its density over the stability
threshold. Thus, isolated gas galaxy disks are likely examples of
self-regulated, non-equilibrium steady states, at least in regions
where the rotation curve is essentially flat (Note that the details of
the self-regulatory processes are not well understood. See
\citet{str99b} for a self-consistent model in the case of strong
global SF. See \citet{zha03} for a discussion of how spiral waves may
be maintained for relatively long periods.)

Collisions upset steady state disks, even if they don't tear regions
in them apart, as occurs in the case of direct collisions between two
gas disks (e.g., \citealt{str97}), or major mergers (e.g.,
\citealt{bar96}, \citealt{mih96}). The waves in these disturbed disks
are of a different nature than those in steady disks. For example,
tidal tails are material, rather than phase waves, and in most cases
induced spirals and bars have mixed material and phase
aspects. Induced waves can have a very different combinations of
Fourier modes than steady waves. For example, odd numbered, asymmetric
modes are evidently more common.

Compressions in steady spirals can push the gas above instability
thresholds and drive SF, but the degree of compression is limited by
the passage time through the wave (e.g., half the epicyclic
period). In material waves, compressed gas elements can move together,
and maintain their compression for longer periods. Beyond this, gas
clouds can be partially separated from their original surroundings,
and launched like collisionless stars over substantial distances, to
interact with other clouds from very different radial positions in the
initial disk. This tidal mixing can sometimes involve substantial
relative velocities, and may play a great role in induced SF. This
point has not been studied in any detail, probably because of the
difficulty in obtaining observational evidence of the mixing.

Tidal mixing is similar to collisional splash effects, where direct
collisions between gas disks drive gas out of both disks, and both
disks experience later fallback. Both effects are analogous to splash
and mixing effects in water waves. Tidal tails are breaking waves in
galaxies.

These examples highlight how detailed studies of SF in colliding
galaxies can advance of understanding of SF processes in general, as
well as allow us to study modes that simply don't occur in
quasi-steady isolated disks. These modes are likely to be very
important in the early stages of galaxy buildup. We will return to the
subject of high redshift galaxies below.

\subsection{Observational Samples of Star Formation Before Merger}
\label{sec:8}

Given these theoretical motivations, let us consider observational
results. In Sec. 1 we discussed the discovery of ULIRG
super-starbursts in gas-rich, major mergers in the 1980s. Generally,
no such strong signal of enhanced SF has been found in pre-merger
interactions. Since in most interactions there is no wholesale gas
compression like that found in major merger remnants this is not
surprising. The questions remain, however, do interaction induced
disturbances lead to substantial SF enhancements, and if so,
where, when and how? These questions were raised by \citet{too72} and
\citet{lar78}. They have been the focus of much interest
in observational studies in many wavebands of both individual systems
and samples of systems (see Sec. 7 in the review of \citealt{str99a}).

The common conclusion was that there is an average SF enhancement in
interacting systems, and that this could be result of a modest
starburst in most cases. However, SF is not obviously enhanced in all
interactions, and may be suppressed in some. The galaxy samples
studied in the 1980s and 1990s were not generally large enough to
provide strong enough statistical results to be definitive, let alone
to tease out details of the relevant processes. The larger samples
tended to contain systems from a wide range of pre-merger or merger
stages (like the the Larson and Tinsley Arp Atlas sample), and so,
could be dominated by the merger-burst effect. On the other hand,
samples of specific types of interaction (e.g., the ring galaxy sample
of \citealt{app87a}) or specific stages (e.g., the \citealt{bus87}
violently interacting sample) tended to be small. Interactions are
rare in the present day, and specific types are therefore doubly rare!

\begin{figure}
\centering
%
\caption{Image of the Arp 89 system (NGC 2648, from \citet{arp66}. It is
an example of systems studied by \citet{kee93}. The companion has one of
the strongest nuclear SFRs in the sample.}
\label{fig:2}
\end{figure}

Nonetheless, interesting clues came out of many of these studies. An
important example is the \citet{kee93} spectroscopic study of SF
correlations in a sample of 75 Karachentsev spiral pairs (see Figure 2
for an example system). This work built on a decade of earlier work by
Keel, Kennicutt and collaborators (\citealt{kee85},
\citealt{ken87}). Keel found that the current SFR (as measured by
H$\alpha$ equivalent width) did not depend much on the projected
separation of the two galaxies, nor on whether a galaxy experienced
the collision as prograde or retrograde. These results seem to defy
the intuitive notion that strong perturbations at closest approach
should drive strong responses, which could result in enhanced SF (but
see \citealt{kee03}). In prograde encounters the companion orbits in the
same sense as the galaxy's spin, and so the encounter is prolonged,
undoubtedly resulting in more disturbance, e.g., tidal tails. Thus, it
was surprising that Keel did not find a spin/orbit effect in the SF.

What Keel did find was SF enhancement in systems with disturbed
kinematics or in galaxies with large regions of solid body
rotation. Disturbed kinematics was measured by the largest difference
between the measured velocity and that of a mean symmetric rotation
curve. Such kinematic disturbances can be seen in barred
galaxies. However, Keel's sample did not include many barred
galaxies. Keel also found that both disk and nuclear SF enhancements
were linked to kinematic disturbance, which at first sight seems to be
another mysterious result.

Keel considered some of the theoretical mechanisms proposed to account
for induced SF in light of his observational results. He found
contradictions between several of the observational results and the
predictions of models on the enhancement of collisions between massive
gas clouds. The correlation of enhanced SF with the size of solid-body
rotation regions lead him to favor gravitational instability
processes, since such regions are very susceptible to these
instabilities.

Recently, Barton and collaborators (\citealt{bar00}, \citealt{bar03})
re-examined these questions with a larger sample of 502 galaxy pairs
and groups drawn from Harvard redshift surveys. In contrast to Keel
they found a significant anti-correlation between SF (again measured
by H$\alpha$ equivalent width) and separation of the galaxies. The two
samples have comparable ranges of separation and equivalent
width. Although Barton et al.'s SF-separation anticorrelation is
statistically strong, it does appear to be strongly influenced by the
approximately two dozen sample galaxies with equivalent widths greater
than or about equal to 50. Given the relative sample size, we would
expect to find only about 3-4 such systems in Keel's sample. Indeed,
there are 4. Thus, it appears that the effect is too weak to have been
easily detected in a sample much smaller than Barton et al.'s. Barton
et al. speculate that the cause of this anticorrelation is driven gas
inflow before merger in some systems.

Barton and collaborators find a second anticorrelation between SF and
line of sight velocity separation between the two galaxies in each
pair. This is in accord with the intuitive notion that slower passages
induce stronger collisional responses. Among their pairs and groups
they find a very strong anticorrelation between SF and galaxy density,
which they interpret as a symptom of the well-known density-morphology
relation in groups and clusters. And finally, \citet{bar03} have
compared their observations and models of H$\alpha$ equivalent width
and B-R broad band color, taking careful account of reddening
effects. They find a significant correlation between burst population
age and separation. They attribute this correlation and post-starburst
spectral indicators in some systems to starburst triggering at closest
approach, and subsequent aging as the galaxies move to apogalacticon.

On the other hand, \citet{ber03} have recently questioned the whole
notion that there are statistically significant SF enhancements before
merger. They examined the UBV broad band colors of a sample of 59
interacting or merging systems, and compared to a control sample of 38
isolated galaxies. They find no significantly greater scatter in the
colors of Arp atlas galaxies relative to controls, in contrast to the
Larson and Tinsley result, and no evidence for a significant
enhancement in global SF in their interacting sample relative to the
control. They do find evidence for a modest enhancement, by a factor
of 2-3, in the central SF of their interacting sample. Given the
previous result this implies a diminution of the average extra-nuclear
disk SF in the interacting sample. \citet{kee93} found no such
distinction between net and nuclear SF enhancement in his sample.

On the face of it, Bergvall et al.'s primary result about the lack of
SF enhancement in interactions seems to contradict many previous
studies. However, these studies also find that the effect is weak if
we exclude merger remnants, and the Barton et al. papers in particular
suggest that we may need a sample of at least several hundred galaxies
to find it. Given Bergvall et al.'s sample size, their work may not
provide strong evidence for the complete absence of an effect, and
they may even be a bit pessimistic in their estimate that the
frequency of strong, triggered starbursts in interacting systems is of
order 0.1\%. Recent very large surveys of galaxy properties, like the
Sloan Digital Sky Survey (SDSS) and the Two Degree Field (2dF) survey,
could provide the answers, and indeed, a couple of analyses based on
these surveys have been published recently.

\begin{figure}
\centering
%
\caption{Specific star formation rates for 3 subsamples of Sloan
Digital Sky Survey galaxies, selected according to absolute SFR in the
ranges: 0-3, 3-10, and $>$ 10 M$_{\odot}$ yr$^{-1}$ (courtesy
B. Nikolic). See \citet{nik04} for details.}
\label{fig:3}
\end{figure}

\citet{nik04} selected nearly 12,500 pair systems with companions
within 300 kpc of the primary from the SDSS. This is a volume-limited,
low redshift sample with SFRs determined from SDSS (extinction and
aperture corrected) H$\alpha$ data, supplemented by IRAS data. They
also reject very close pairs, i.e., most merger remnants. They find
that ``the mean projected star formation rate is significantly
enhanced for projected separations less than 30 kpc.'' (see Figure
3). Like Barton et al. they find an anticorrelation between SF and the
pair velocity difference. Despite its statistical significance they
also found the the SF-separation anticorrelation is relatively weak,
in accord with previous studies.

With such a large sample, they were able to look at subsamples, for
example, subsamples consisting of two late-type disks, two early-type
disks, or mixtures. The anticorrelation is present in all three
subsamples, with some indication that it extends to larger radii in
the late-type subsample. Nikolic, Cullen and Alexander also presented
SFRs normalized by galaxy mass, and show that the magnitude of the
normalized SF-separation relation depends on how the normalization is
performed.

\citet{lam03} carried out a similar pair study with 1258 pairs from
the 2dF survey, and found anticorrelations of SF with separation and
velocity like those in the Nikolic et al. study.

Bergvall, Laurikainen, and Aalto noted that ``the interacting and in
particular the merging galaxies are characterized by increased far
infrared luminosities and temperatures that weakly correlate with the
central activity.'' This result, in turn, agrees with much evidence
that many specific types of interacting galaxy have enhanced
far-infrared emission. For example, M51 type systems (e.g.,
\citealt{kli01}) on one hand, and the collisional ring galaxies
\citep{app87a} on the other hand, both show modestly enhanced IRAS fluxes
relative to the late-type disk norms.

\begin{figure}
\centering
%
\caption{Some examples of LIRG systems in B and I wavebands. Note the
scale bars and the change of scale between rows. From \citet{arr04},
courtesy S. Arribas.}
\label{fig:4}
\end{figure}

For a broader perspective, one can turn the table and ask about the
nature of galaxies with enhanced infrared emission (and usually radio
continuum emission as well). We have discussed ULIRGs above, and noted
that they are primarily merger remnants, and so not of interest in the
present context. Luminous Infrared Galaxies (LIRGs or LIGs) and Very
Luminous Infrared Galaxies (VLIRGs or VLIGs) are variously defined as
galaxies with far-infrared luminosities in the approximate range 3 x
$10^{10} - 10^{12}$ L$_{\odot}$, and have not been studied as intensively
(see examples in Figure 4). However, it appears that a large fraction
of these objects are pre-merger, collisional systems with a relatively
strong starburst in the core of at least one of the galaxies (see
e.g., \citet{you96}, \citet{gao99}, \citet{cor03}, \citet{arr04} and
references therein). Based on statements like that of Bergvall et
al. in the previous paragraph, and the rarity of LIRGs (like the
ULIRGs), it seems likely that they are the same as, or more obscured
relatives of, the few starburst galaxies that seem to be responsible
for the weak SF enhancement found in optical pair samples.

In sum, optical studies show that interactions lead to only a very
small SF enhancement before merger, on average. Given that core
starbursts are likely to have quite short durations (unless they have
prolonged driving, e.g., \citealt{str05}), it is natural to interpret
this as the result of random sampling of a common process with a short
duty cycle. The LIRG studies suggest another possibility, that a small
minority of galaxies (the LIRGs) are responsible for the general weak
enhancement, and that these galaxies are near the end of the road to
merger, though not yet merged. The latter clause is supported by the
fact that the few starbursts in pair samples generally have small
separations and velocity separations, and this is also true for many
LIRGs. In this alternative view, SF is not significantly enhanced in
the early stages of interaction despite strong morphological
disturbances. Also there is a more continuous increase in SF as merger
is approached, an idea suggested in some of the LIRG studies. Gao and
Solomon, in particular, have suggested that the phase structure of
galactic ISM changes through the merger process, with an increase in
the molecular phase in the final pre-merger stage (\citealt{gao99},
also see \citealt{gao04} for similar results concerning molecular
abundance changes through core starburst evolution). We will return to
this discussion in Section 2.5.

\subsection{Detailed Case Studies}
\label{sec:9}

Because it is difficult to directly translate projected galaxy
separations into true separations, and directly divine the stage along
the path to merging, it is difficult to use limited observations (in
any waveband) to determine which of the viewpoints of the previous
paragraph is correct. (Although it might be possible to estimate the
separation and evolutionary stage statistically in the large samples.)
There are two other ways to test evolutionary hypotheses. The first is
to confront it with theory and the results of numerical simulations,
which we will consider below. The second is by assembling a large
library of careful case studies of specific systems. Such studies
require a panchromatic array of spatially resolved and kinematic
observations, which can provide strong constraints on numerical
models. They also require system specific models, which closely match
all available observations, and thereby provide a clear determination
of the interaction stage (see discussion in Struck 2004). Given the
prolonged debate on whether the nearby M51 system is the result of one
or two close encounters, this is not necessarily an easy task (see
review of \citealt{str99a}), though in either case it is clear that
the system is not yet near the end of the merger road.

With a library of detailed case studies one could hope to graph SFR
(or specific SFR per unit mass or gas mass) versus interaction stage
to resolve the issues above. The ``interaction stage'' would require
careful definition, however.

Detailed color and spectral synthesis modeling can in fact yield
constraints, if not yet unique solutions, for the SF history of some
nearby well-studied systems, e.g., the Magellanic Clouds
(\citealt{zar04}, \citealt{jav05}), M51 (\citealt{bas05},
\citealt{bia05}), M82 (\citealt{deg01}, \citealt{deg02}) and Arp 284
\citep{lan01}. From such cases, one can add a few points to the
hypothetical SFR-interaction stage plot.

\subsection{Modes of Star Formation}
\label{sec:10}

Detailed case studies are also a primary tool for studying a number of
specialized modes of SF, some of which have received a good deal of
attention in recent years. These include: the formation of super star
clusters (SSCs), SF in tidal bridges and tails, and SF in induced disk
waves. Except perhaps in the last case, these modes do not usually
dominate the SF in interacting systems, but they may involve physical
processes unique to collisional environments, and produce especially
interesting products like dwarf galaxies and halo globular clusters.

\subsubsection{Tidal Dwarf Galaxy Formation}
\label{sec:11}

We have already mentioned early studies of SF in tidal tails in
Section 1, but there has been a great of recent work. Work in this
area has been energized by the possibility that, not only massive star
clusters, but actual dwarf galaxies might be formed out of material in
tidal tails (Figure 5). If so, this could be a means of forming dwarf
galaxies at the present time, and in observable environments. In the
introduction to a recent paper \citet{duc04a} review much of the
literature of the last decade, and additionally a section of the
proceedings of a recent IAU symposium is also dedicated to the topic
\citep{duc04b}. These two sources provide a good entry points to the
literature.

\begin{figure}
\centering
%
\caption{Arp atlas image (Arp 1966) of the Arp 105 system. This system
contains a probable tidal dwarf galaxy at the end of the long tidal
tail in the north. See \citet{duc97} for details.}
\label{fig:5}
\end{figure}

We should begin by noting that the tidal dwarf formation has been
controversial, and difficult to prove (or disprove). Most tailed
galaxies do not have an obvious luminous SF region at the end of their
tails. To date, only a few examples of dwarfs forming in tails have
been studied in detail, so it is not clear how rare is that
circumstance, nor what is the general nature of SF in tails. In fact,
there are a number of difficulties in finding these objects, and
confirming that they are dwarf galaxies in formation. Sometimes the
tail is viewed edge-on, and if it is curved in the vicinity of the
candidate dwarf a good deal of material that is not physically
connected can be superposed along the line-of-sight, including
multiple SF regions (e.g., \citealt{duc00}). This can lead to large
overestimates of the mass and extent of the tidal dwarf candidate,
leading, in turn, to a bias for such systems in the candidate
list. Determining whether a tidal dwarf candidate is truly a
gravitationally bound object, and will persist as a distinct entity is
also challenging, if only because of the resolution limits of
observations of HI and molecular gas in these small objects.

Discussions of this latter question have been entwined with those
concerning two early theories for the formation of tidal
dwarfs. \citet{bar92} suggested that they could form as a result of
gravitational instabilities in tails consisting of collisionless
stars, while \citet{elm91} argued for the dominance of gas dynamical
processes in regions of enhanced turbulence (i.e., enhanced velocity
dispersion). One of the difficulties in the modeling is that some
density concentrations may not be persistent, and the models are not
generally able to follow their evolution for very long times, or with
sufficient particle resolution (but see the high resolution model of
\citealt{hib04}). Another problem in confronting these theories to
observation is that since gas disks are more extensive than stellar
disks, all tidal dwarf candidates are likely to contain a large
fraction of gas, so it is not possible to find a case of assembly by
gravitational means alone. Based on new simulations, \citet[and
references therein]{duc04a} argue that only if the parent galaxy has
an extensive dark halo is it likely that large amounts of gas will
accumulate at the end of a tidal tail, and that this is the most
efficient route to forming true tidal dwarfs with masses in excess of
$10^9 M_{\odot}$. These authors also find that the gas accumulation
process is primarily kinematic, with self-gravity playing only a minor
role. It will be interesting to see how these new ideas develop in the
next few years.

\subsubsection{Massive and Super Star Cluster Formation}
\label{sec:12}

\begin{figure}
\centering
%
\caption{HST image of young star clusters in the merging Antennae
galaxies, from \citet[ courtesy of B. Whitmore]{whi99}.}
\label{fig:6}
\end{figure}

One of the greatest contributions of the Hubble Space Telescope to
extragalactic astronomy was to resolve individual star clusters in
relatively nearby galaxies and allow us to take the census the cluster
populations in them. As a result of such studies it has become clear
that a large fraction of new stars in colliding galaxies are formed in
clusters (see Figure 6). This is difficult to quantify, but has been
estimated at 50-100\%. The characteristics of the most massive of these
clusters, the super star clusters with estimated masses in the range
$10^5 - 10^8 M_{\odot}$, are just what we would expect from young clusters, so
it appears that we are now able to study the formation and development
of globular clusters at a variety of stages by direct
observation. These studies have given rise to a considerable
literature, which extends far beyond the topic of this review, so we
only describe a few of the relevant highlights.

In a summary of a recent conference on this topic, \citet{oco04}
emphasized the universality of the properties of young cluster
populations, despite a huge range of formation environments and
scales. These properties include a nearly universal power law mass
function, which evolves naturally with time to the exponential
function of old globulars. The number of clusters and the maximum
cluster luminosity in a star-forming region both scale with total
SFR. Most cluster populations have a very small range of formation
ages. This is especially true of populations in galactic nuclei, but
in colliding galaxies with widely spread SF regions there can be
distinct populations, each with small age spreads (e.g.,
\citealt{alo02}). The spatial structure of super star clusters also
appears to be universal. The stellar initial mass function is
universal, at least at the high end where it can be determined.

It is worth emphasizing the range of environments where super star
clusters and their somewhat less massive relatives are found in
colliding galaxies. Of course, starburst nuclei are primary locations
and M82 (\citealt{deg01}, \citealt{mel05}) is probably the most famous
example. M51 (\citealt{bik03}, \citealt{bas05}) is also very
interesting. At the other extreme we have globular cluster populations
of intermediate age (i.e., of order a few Gyr) around merger
remnants. In the ongoing merger in the Antennae system, clusters are
scattered at many locations in the bodies of the galaxies
\citep{whi95}. Massive young clusters are found in many tidal tails,
though interestingly \citet{kni03} make a suggestion, based on their
study of 6 tail regions, that they either have a population of massive
young clusters or a tidal dwarf, but not both. This conjecture
certainly merits further observational and modeling study. Massive
cluster populations are also found in ring galaxies like the Cartwheel
\citep{app96b} and ocular waves like IC 2163 \citep{elm00}. It seems
very likely that the mid-infrared detectors on the Spitzer Space
Telescope will find massive cluster populations in more environments
that are hidden from Hubble and ground-based telescopes.

What do these environments have in common and what's the physics
behind massive cluster formation? \citet{oco04} summarizes the
prevailing view that the formation of SSCs requires high gas
pressures, of order $10^4$ times those of the interstellar medium in
the solar neighborhood, and that these high pressures must extend
through a region of size greater than 1 kpc (also see
\citealt{schw98}). Strong turbulence also pervades the formation
region. O'Connell emphasize that the energetic environment inside a
forming massive cluster must be truly extraordinary.

It is likely that all of the colliding galaxy environments noted above
are able to achieve the high pressures and turbulence that the theory
says are necessary to form the super star clusters. This is not
entirely clear in the case of disk waves and tidal tails. However, in
the former case the process may be aided by feedback effects from the
first stars to form. In the case of tails it may simply be that some
achieve the requisite conditions and form massive clusters, and others
do not. We have much to learn yet about these processes.

Finally, O'Connell notes a couple examples of nuclear starbursts where
the super star clusters are much more massive than the other clusters,
and so, the mass function is discontinuous. He speculates there may be
a special formation mode for these cases, though the nature of that
mode is not clear. As in the case of tidal dwarf formation there are
competing mechanisms, and one of these may dominate only in the
exceptional cases. These mechanisms again include the formation of
massive progenitor clouds triggered directly via gravitational
instability, or indirectly in dense environments assembled by large
scale gravitational instability. They may also include hydrodynamic
effects like cloud crushing that occurs when giant clouds experience
an abrupt pressure increase after impacting large-scale shocks or
other high pressure regions (\citealt{jog92}, \citealt{bra04},
\citealt{bek04}). The combination of these processes could probably
generate discontinuous cluster mass functions, but at present, this is
only speculation.

A few more exotic ideas have also been discussed
recently. \citet{sca04} have suggested that strong winds from young
galaxies could have shocked their dwarf companions, stripping gas and
compressing it to form globular clusters. \citet{bur04} also suggest
that globulars might form in dwarf companions. Their argument is based
on a hierarchical clustering model of galaxy formation as applied to
the cluster populations in the Milky Way and M31. On the other hand,
\citet{hib00} have found several examples of systems where winds seem
to have swept the gas out of parts of tidal tails, without the
production of massive star clusters.

\subsubsection{ULXs}
\label{sec:13}

Ultraluminous X-ray sources (ULXs) are defined as having X-ray
luminosities of order $10^{39} - 10^{41}$ ergs per s$^{-1}$, which
extends beyond the luminosities of the well-studied high mass X-ray
binaries, but is still much less than a typical active galactic
nucleus. X-ray sources of this luminosity have been detected in
galactic nuclei for two decades, but the arcsecond resolution of the
Chandra Observatory has facilitated their discovery and definition as
a class of objects (see reviews of \citealt{mus04a},
\citealt{van04}, \citealt{war03}).

Estimates indicate that they may be a quite common constituent of the
nuclei of normal galaxies. However, their nature remains somewhat
mysterious. There are two leading theories. The first is that they are
indeed an extension of the high mass ($\simeq 10 M_{\odot}$) binary
phenomenon, but with highly super-Eddington accretion rates and beamed
emission (e.g., \citealt{beg02}, \citealt{kin04}, \citealt{kin05},
\citealt{liu05}). The second is that these are in fact black hole
accretion systems of intermediate mass between stellar black holes and
active nuclei, e.g., masses $>>$ 100 M$_{\odot}$ (e.g., \citealt{hop04},
\citealt{kro04}, \citealt{mil04a}, \citealt{mil05}). There are strong
arguments for both models. For the most luminous ULXs the stellar mass
explanation is strained. For the most rapidly variable ones the source
size and mass is limited from above. Since the luminosity bounds of
the class are ad hoc, it is certainly possible that the class includes
both kinds of source. \citet{gut05} also present a cautionary tale of
how apparent ultraluminous sources can in fact be background sources.

\begin{figure}
\centering
%
\caption{Smoothed greyscale X-ray image of the Arp 284 system from
Chandra Observatory data, with luminous point sources
numbered. Contours showing the optical outline are derived from the
Digital Sky Survey image. See \citet{smi05} for details.}
\label{fig:7}
\end{figure}

A number of galaxies are now known to contain enough ULXs to allow the
construction of luminosity functions, and these luminosity functions
are found to scale with total SFR or a combination of SFR and galaxy
mass (e.g., \citealt{swa04}). Over the last 5 years ULX populations
have been found in several colliding galaxies, including M82
(\citealt{fio04}, \citealt{mat04}, \citealt{por04} and references
therein), the Antennae, the Cartwheel ring galaxy, and the ring-tailed
Arp 284 system (Figure 7). The ULXs are widely spread across the
Antennae (\citealt{fab04}, \citealt{mil04b}. They are found in the
outer ring and an X-ray bridge of the Cartwheel (\citealt{gao03},
\citealt{wol04}). In Arp 284, most are associated with tidal features,
especially a prominent tail, but with one of the brightest known
contained in the ring (\citealt{smi05}). Some of these
ULXs are associated with star-forming regions, and so they provide
another locator and probe of such regions that is not easily obscured.

The environment of these ``tidal ULXs'' can be less confused than that
of many nuclear ULXs. A few are associated with super star
clusters. Models can often provide a good deal of information about
the dynamical history of specific tidal structures, and thus, about
the formation environment of the ULXs. The discovery of any specific
phenomenology in these environments could provide useful information
on the nature of ULXs. However, at present the number of cases studied
is too small to allow any firm conclusions beyond the association
between collision morphologies and the occurence of ULXs.

\subsection{Summary and Theory}
\label{sec:14}

Clearly, there has been a vast amount of work on interaction induced
SF in the last few decades, and we have only been able to skim through
it above. Is all of this work leading to a comprehensive understanding
of the phenomenon? Perhaps not quite yet, but we may be getting close
to that goal. The question can be broken down into several separate
questions. First of all, do we understand the general physical
processes, and do we understand enough about how they work to account
for the general observational systematics? Secondly, do we understand
these processes well enough to reproduce their effects in numerical
models, both comprehensive models of specific systems, and models of
SF in particular dynamical processes?

\subsubsection{Numerical Models}
\label{sec:15}

We'll begin by considering recent numerical models and some aspects of
the last question. Star and star cluster formation take place on
scales that are orders of magnitude smaller than those typically
resolved in simulations of galaxy collisions. However, separate models
of the process on those small scales are beginning to advance our
understanding greatly. Because of this, and the fact that much of the
dynamics on the intermediate scales is essentially scale-free
turbulence, we may be able to develop reasonably accurate SF
formulations, without needing to resolve the scales on which it
occurs. However, to date, relatively simple SF prescriptions have been
used in galaxy collision models. Moreover, these prescriptions have
been based on several different ideas about the dominant SF triggering
process. Three of the most popular are: 1) a simple density-dependent
SFR, 2) triggering by strong compressions in cloud-cloud collisions or
large-scale shock waves, or 3) triggering by gravitational instability
above a threshold density (or surface density or pressure).

With regard to the first of these, the Schmidt Law, in which the SFR
is proportional to a low power of the gas density, surface density, or
the gas density divided by a local dynamical time has been surprising
resilient. \citet{mih93} used it (and isothermal particle
hydrodynamics) to model collisions between two disk galaxies, and
\citet{mih94} used it to model the Cartwheel ring galaxy. In both
cases they found that the models gave about the ``relative intensity
and morphology of induced star formation.''  Later, \citet{mih96}
found that with this formulation mergers between disk galaxies with
bulges could produce burst SFRs a hundred times larger than those of
isolated galaxies. They also explored how the presence of a bulge
component affected the merger SFR. Given the simplicity of the
prescription the results are impressive. However, \citet{cox05}
have recently shown that the amount of SF in mergers may have been
overestimated in earlier models, because this quantity depends on how
conservation conditions are implemented in the SPH algorithm.

Phenomenological cloud collision models for gaseous dissipation in
galaxies go back to the 1970s. The obvious disadvantage of such models
is that interstellar gas clouds are transient, ever-changing
structures, and not the coherent entities implicitly assumed when
equating them with the 'sticky' (i.e., dissipative) particles of a
numerical model. On the other hand, it is a straightforward way to
model the cloud collisions and shock encounters which occur in many
types of collision. In models of polar ring galaxies \citep{bek97},
models of starbursts in multiple mergers \citep{bek01}, and in other
applications, Bekki has used a hybrid particle model. In his models
there is dissipation from cloud collisions, but a probabilistic
Schmidt Law is used to convert selected gas particles to stars. The
local gas density is computed for each gas particle and used in the
Schmidt Law.

Recently, \citet{bar04} has proposed a rather sophisticated
phenomenological model, in which SF depends on the amount of energy
dissipation in shocks. He argues that with this prescription he is
able to produce a much better model of the Mice system (NGC 4676) than
with a Schmidt Law. This is one of the few significant comparisons of
different formulations in models of the same system. Barnes also notes
that Schmidt Law models are quite insensitive to details of the
interaction, while shock induced SF is very sensitive, and could be
checked observationally.

Threshold instability models have been used frequently in the areas of
galaxy formation and multiphase models of galaxy disks in recent
years. This author has used such a model with feedback and gas with a
continuous range of thermal phases in studies of direct collisions
between two gas disks and their reformation \citep{str97}. More recent
work on disk collisions with many more particles has been carried out
by \citet{spr05}. \citet{cox05} have recently presented an efficient
effective equation of state approach to handling the thermal physics.

I have also used this type of SF formulation in detailed N-body
hydrodynamic models of a couple of specific systems with extensive
observational data (\citealt{str03}, \citealt{str05}). Both the
spatial distribution of SF and the history of net SF fit the
observational constraints, though the constraints on the SF history
are not stringent. At low threshold densities this type of formulation
is probably much like the Schmidt Law, since the SF will occur in
regions with the most particles (i.e., high density). With a high
threshold density, a violent process like shock compression and
subsequent cooling will be needed in many cases to drive SF, more like
the Barnes model.

In the end we see that many different numerical treatments can
simulate induced SF reasonably well, and so none are immediately
falsifiable. The answer to the question posed at the beginning of this
discussion is yes, we can reproduce observations, but not because the
models represent the underlying physical processes especially
faithfully. The universality of those processes, and their highly
interconnected properties, allow modelers to use simple formulations
on large scales. Stringent tests of feedback prescriptions may
eventually come by fitting the mass fraction and distribution of
warm-to-hot phases in the interstellar gas. However, this will take
much more realistic modeling of the thermal physics, and the stellar
initial mass function, than is currently the norm.

Detailed observational studies on kiloparsec scales in various
environments may provide insight into how sensitive SF is to
compression and dynamical timescales. Spitzer Space Telescope
observations in the mid-infrared have the ability to see through
obscuring dust and provide a complete SF census on these scales in
nearby galaxies, so the prospects are exciting in the next few years.

\subsubsection{Squeezing Out Stars}
\label{sec:16}

All of the models described above, form stars by compressing gas
(albeit in more or less finely tuned ways). This recalls the
\citet{ken89} observational result on the universality of the Schmidt
Law over a range from isolated galaxies with modest SFRs to
ULIRGs. Apparently, the first law of induced SF is - it's just the
(large-scale) compression. More precisely, it appears that large-scale
compression drives a turbulent cascade, which enhances star-forming
compressions on the small scales (e.g., \citealt{kru05}, and
references therein). Because of the universality of the cascade, this
process doesn't necessarily depend much on the details at the large
and small scales.

In the case of ULIRGs the spectacular response is the result of
spectacular angular momentum transport and compression in the major
merger. For rapid or distant encounters the most that can be achieved
are relatively small compressions in bars and waves. It is worth
recalling that basic tidal forces stretch along the line of galaxy
centers and compress in the perpendicular directions. For an
approximately two-dimensional disk this means (very roughly!)
stretching in one dimension and compressing in one
dimension. Alternately, in terms of a simple impulsive torque, it
means that angular momentum is added to one side of the disk (stars
are pulled ahead in their orbits), and subtracted from the other side
(stars are pulled back). Thus, net compression across the disk is
roughly balanced by stretching or torque-induced rarefaction. In
either case, the global effects are modest for small amplitude
disturbances, implying little induced SF, as observed in such
cases. On the other hand, in strong disturbances, the torque-induced
decelerations of the gas orbital motion, and subsequent compression of
a significant fraction of the gas, may be enough to induce a strong
starburst, regardless of the fate of the rest of the gas.

LIRGs seem to be a heterogeneous class, but they include interacting
galaxies that are separated by about 1-2 diameters of the larger. In
such cases the tidal effects are nonlinear. In addition, the
gravitational forces within each disk will be augmented by dark matter
from the other galaxy's halo, which is coextensive with the disk. (In
fact, the importance of this effect must be estimated quantitatively,
but generally it will become important at the separations cited.) The
resulting global compressions can account for the SF enhancement. It
appears that such cases play an important part in creating the
observed anti-correlation between SFR and the separation of the two
galaxies.

In sum, it appears that the general systematics of induced SF can
indeed be accounted for, to first order, as the direct result of
compression. The consequence of this, that we can learn little more
about SF physics from large scale studies, is disappointing. On the
other hand, it means that colliding galaxy model results are not
sensitive to many details of the SF/feedback formulation, and that
there is little point in trying to extend numerical resolutions to
very small scales. (However, we will eventually have much higher
observational and modeling resolutions, which would allow the study of
the cloud turbulent cascade, and the full effects of the
non-equilibrium interaction environment. The point is that modest
resolution improvements will not help much.) 

We conclude with a brief mention of some possible exceptions or
refinements to the ``it's just compression'' rule. The first might be
found in the environment of core starbursts. If these are triggered by
strong shocks, or with a sensitive threshold, then they may turn on
rapidly. If, instead, they obey the Schmidt Law they will turn on more
slowly if the central gas mass and density accumulate
slowly. Generally, current spectral synthesis techniques are not able
to provide SF histories that are accurate enough to distinguish. That
is, except in a few nearby starburst cores, where the evidence seems
to favor rapid turn-on of local density concentrations (because of the
small age spreads within these concentrations, e.g.,
\citealt{har04}). Similarly, the study of core burst turnoff might be
enlightening. There seem to be regions in the core of M82 where the
cloud system is disrupted, SF is turned off, but pressures and gas
densities remain high (\citealt{mao00} and references therein). The
existence of such regions may necessitate at least a caveat in the
Schmidt formulation.

We might be able to derive more information by studying SF within and
behind density waves in disks. Spiral waves are ubiquitous, but ring
waves produced in direct galaxy collisions are simpler. Asymmetric
rings produced in slightly off-center collisions are the most
interesting because they are still simple, but the wave amplitude
varies continuously with azimuth. This is an excellent environment for
confronting threshold versus continuous theories, at least if the
compression in part, but not all, of the ring exceeds the
threshold. To date, there is some evidence in support of thresholds,
but not without complications ranging from incompletely known
obscuration to unknown details of the collision parameters. The
primary example is the Cartwheel ring with large variations of SFR and
cluster populations around the ring, but with uncertainty about the
details of the collision, and no old star component in the ring to
provide independent information about wave amplitude as a function of
azimuth (\citealt{app96a}). With its relatively high resolution and
ability to see through much obscuration, the Spitzer space telescope
could resolve these ambiguities in carefully chosen systems.

There are many more examples of how to get beyond the simple
compression law and the first-order theory of induced SF, and such
work should become increasingly important in this field.

\section{Environmental Effects}
\label{sec:17}

We have seen that the study of galaxy collisions is a relatively
young, but rapidly maturing field. Thus, it is understandable that
most progress to date has been in understanding the most spectacular
collisions, ULIRG/major mergers, and the nature of some of the closest
systems which can be studied in detail. Most of the latter occur in
quite small groups like our own local group. However, studies of
collisions in other environments date back to \citet{spi51}, and their
number has been increasing recently.

\subsection{Cluster Bustle}
\label{sec:18}

\begin{figure}
\centering
%
\caption{\citet{arp66} atlas image of Arp 120 (NGC 4438), a Virgo
cluster system in collision, with a starburst galactic wind, and
likely also experiencing environmental effects. See \citet{bos05} and
\citet{ken02} for details.}
\label{fig:8}
\end{figure}

At the opposite end of the spectrum from the local group environment
is that of massive, dense clusters of galaxies. It will suffice for
our purposes to briefly note the different processes in this
environment relative to that of local groups. These differences
include: high speed collisions, galaxy 'harassment,', ram pressure
stripping, 'strangulation,' and induced slow collisions (see Figure
8).

\citet{spi51} first suggested that high velocity collisions in galaxy
clusters might have little effect on the stellar components, but could
blast away the overlapping parts of gas disks. This is because the
typical random galaxy velocities in clusters of up to several thousand
km/s, are not only highly supersonic for the intercluster gas, but are
in excess of normal disk escape velocities. Generally, we expect a
moderation of tidal gravitational effects, but in some cases a drastic
increase in hydrodynamic effects.

High-speed collisions may have weak gravitational effects, but
encounters are much more common in the cluster environment. Thus, in
the aggregate, tidal effects are not negligible in clusters. The
cumulative effect of many weak (high-speed or distant) galaxy-galaxy
interactions in clusters, as well as perturbations from the cluster
potential, and possibly from intermediate scale sub-structure is
called 'harassment' (see \citealt{moo96}, \citealt{moo99}). In
recent high resolution numerical studies of the growth of moderate
clusters, \citet{gne03a, gne03b} has shown how this process can secularly
erode galaxy halos, thicken moderate mass stellar disks and truncate
SF, and destroy small disks.

Ram pressure stripping (RPS) by the intra-cluster medium can have
somewhat similar effects on disks. RPS is an interesting subject, with
a number of recent developments, and worthy of a review of its
own. Thus, it is beyond the scope of this review, except for a few
comments. First of all, long time residents of dense galaxy clusters
were probably stripped long ago, so RPS is most relevant to gas-rich
galaxies falling into the intra-cluster medium for the first
time. X-ray satellites have provided much evidence that the infall of
galaxy groups into clusters and cluster-cluster mergers are still
common events \citep{mus04b}. Once the intra-group medium is
stripped in such interactions, individual disk galaxies are vulnerable
to RPS. RPS of spheroidal galaxies has been studied for 30 yrs., but
in the last 10 yrs. a small literature on stripping of disk galaxies
has blossomed.

The outermost parts of gas disks are stripped promptly, and slower
interactions continue for some time. Slow viscous interactions at the
edge of disks moving face-on into the intra-cluster medium has
recently been studied in detail by \citet{roe05}. The
three-dimensional dynamics in tilted cases have been modeled in
several recent papers (\citealt{aba99}, \citealt{vol00},
\citealt{vol01}, \citealt{sch01}). Schulz and Struck, in particular,
pointed out that if the gas disk is not promptly stripped, it can
nonetheless be displaced relative to the stellar disk and the halo
center. The displaced gas disk experiences tidal compression
(perpendicular to the disk plane) and asymmetric torques in the tilted
case, which generate spiral waves. The waves transfer some gas and
much angular momentum outwards, where it is stripped. The remaining
gas, with less angular momentum, compresses radially, which
``anneals'' it against further stripping. The various compressions
should induce SF. The tidal forces and induced SF are much like those
in galaxy collisions. Thus, after the stripped material is gone, it
could be difficult to discern whether excess SF is the result of RPS,
a minor merger, or harassment. In fact, since these processes could
work simultaneously, the question may be academic.

Vollmer and Schulz and Struck emphasized another aspect of stripping:
some of the removed material can later fall back onto the disk. This
can occur either when the galaxy moves into regions of lower
intra-cluster medium densities, where the levitating pressure is
reduced, or when gas clouds move into the disk 'shadow' where the
pressure is also reduced. In either case we would expect effects akin
to those of mass transfer in galaxy collisions.

Strangulation is a weaker cousin of RPS. It is the process of removing
the potential feedstock of disk SF, gas in the galactic halo, usually
via RPS \citep{lar80}. The feedstock could include gas blown out of
the disk by supernovae or stellar or galactic winds, it could include
gas tidally removed from companions, or primordial gas falling into
the halo for the first time. This process is likely to be most
important in the young universe, when there is still much gas outside
of galaxy disks, but it also hampers gas recycling from dying
populations in cluster galaxies.

\subsection{Cluster Slow Dance}
\label{sec:19}

A final process that may be very important in cluster environments is
induced slow encounters in infalling groups. Recent studies of the
Butchler-Oemler effect (see review of \citealt{pim03}), which is an
excess of blue galaxies in clusters at redshifts of less than 1,
provide evidence that many of the blue galaxies are mergers or
interacting (also see \citealt{zab96}, \citealt{has00},
\citealt{ell01}, \citealt{got05}). It seems unlikely that high speed
interactions could be responsible for this effect. \citet{mih04} has
emphasized that large-scale cluster formation models show that slow
interactions continue to occur even in large clusters, and are quite
common during cluster formation. He also notes that slow encounters
can occur in groups or small clusters with modest velocity dispersions
falling into large clusters (also see \citealt{pog04}).

This is an interesting phenomenon that has not been much
investigated. Mihos notes that a number of different processes may be
involved and it may be impossible to disentangle them. For example,
tidal forces from the cluster potential could perturb orbits in the
infalling group inducing interactions, and intracluster medium
annealing could induce increased SFRs. Personally, I suspect these are
secondary processes.

At least for galaxies that fall through the cluster core the primary
process may well be gravitational shocking, which depends directly on
the cluster potential rather than indirectly or on the derivative of
the potential (tidal forces). Consider the relatively simple example
of a galaxy group containing about 30-100 galaxies, falling into a
large dense cluster. Cold dark matter structure formation models
predict a common density profile across the range of structures from
galaxy halos to the dark halos of large clusters, and observational
tracers (e.g., intracluster starlight, see \citealt{fel02} and
references therein) show good agreement with profile functions derived
from these simulations, like the popular NFW profile
\citep{nav97}. Then it is reasonable to assume our example group and
cluster have similar density profiles, though not necessarily with the
central cusp of the NFW profile.

The observations also suggest that the central density decreases
slowly with mass in dark matter halos. We might, for example, model
our large cluster after Abell 2029, whose mass profile was studied in
detail by \citet{lew03}. They find a mass of about 9.2 x
10$^{13}/h_{70}\ M_{\odot}$ contained within a radius of 260/$h_{70}$
kpc, yielding a central density of 0.0052 M$_{\odot}$/pc$^3$. As an
example of a group, on the other hand, we can take a 'poor' group like
those studied by \citet{zab98}. These groups have virial masses of
about 7 x 10$^{13}/h_{70}$ M$_{\odot}$ within radii of about
300/$h_{70}$ Mpc. The authors estimate that 80-90\% of the virial mass
is in the group halo, so the mean group halo density is about 0.0022
M$_{\odot}$/pc$^3$. These numbers are meant to be representative, not
precise. Group parameters, e.g., in compact versus poor groups, could
easily range over factors of a few. Nonetheless, the message is clear
that passing through the cluster core would substantially increase the
instantaneous group halo mass.

Moreover, the time to pass through the core is of the same order as
the group dynamical time. We can estimate the former by dividing the
core radius above by a typical cluster velocity of about 2000 km/s;
the result is about 300 Myr. The free fall time at the edge of our
example group is about 200 Myr. Therefore there is time for group
galaxies to be pulled into a much denser and compact
configuration. For roughly comparable group and cluster halo core
densities, galaxies could be pulled in to roughly half their previous
distance from the core, increasing the galaxy density by nearly an
order of magnitude and their collisions by about a factor of 100
(density squared).

Like stars in clusters that pass through the galactic disk, when the
group leaves the cluster core, and gravity is reduced, the galaxies
will fly outward. (This is also like collisional ring galaxies.)
However, collisions between galaxy halos are stickier than those
between cluster stars. During the compression period, along near
parallel but converging trajectories on the way down, encounters are
likely, and there is time for dynamical friction to dissipate relative
orbital energy. The result would frequently be a fairly slow
interaction and eventual merger.

Considering all the various ways that clusters accelerate galaxy
evolution, one can view life for average galaxies in small local
groups as more or less a holding pattern, or at least a matter of slow
maturation. Evolution doesn't begin in earnest until they fall in a
larger group or cluster; the classic tale of youth leaving the farm
for the big city.

\section{Interactions and Galaxy Evolution}
\label{sec:20}

Galaxy collisions drive galaxy evolution, but how much compared to
other processes like the passive conversion of gas to stars in
isolated galaxy disks or compared to other dynamical processes like
ram pressure stripping?  Also, how does the role of collisions change
with cosmological epoch?

To begin, we note that it has become quite clear that at least some
massive galaxies and massive disk galaxies, in particular, formed very
early, and had already attained a respectable age by redshifts of 1-2
(see review of \citealt{spi04}). From a practical point of view, this
means that observations must push to very high redshifts to see big
evolutionary changes. We will come back to what has been seen in a
moment. This fact has also been taken as evidence that at least some
galaxies formed in a rapid monolithic collapse, rather than building
up steadily in a prolonged sequence of mergers.

\subsection{Models of Structure Buildup}
\label{sec:21}

What do theory and cosmological structure formation simulations lead
us to expect? Currently, hierarchical build-up, $\Lambda$CDM models
(i.e., models with cold dark matter plus ``$\Lambda$'' dark energy) in the
``concordance cosmology'' are the dominant paradigm. This picture
suggests the occurrence of many mergers of small building blocks in
the earliest stages, and continuing mergers thereafter. Moreover,
recent analysis shows that it is possible to form some massive
galaxies, including disk galaxies, at early times in these models
\citep{nag05}.

Thus, part of the solution to the paradox of early massive galaxy
formation is that fully nonlinear $\Lambda$CDM models do not yield
exactly the same results as simple, analytic hierarchical structure
formation models. Early massive disk galaxies may be a roughly
2$\sigma$ outcome of the simulations, but that may be sufficient to
account for the observations.

From another point of view, the early formation of massive disks
allows for the possibility that major mergers form elliptical galaxies
at early times, i.e., accounts for ellipticals containing only old
stellar populations within the merger theory (see discussion in
\citealt{schw05}).

To return to the $\Lambda$CDM paradigm, another thing the simulations
show us is that when substantial entities merge, not all their
substructure is erased. In fact, the absence of hundreds of dwarf
satellites around the Milky Way has been cited as a problem for this
kind of model (e.g., \citealt{kly99}). However, RPS in the hot
halo, tidal disruption, and collisions with the galactic disk may have
destroyed many of the leftover building blocks. Indeed, digesting the
leftovers may be an important secondary evolutionary process,
operating alongside the primary hierarchical merging process.

Thus, the picture of sequential buildup of galaxies via successive
major mergers in the simplest hierarchical models is not the only one
in which collisions and mergers are crucial. In more realistic models
minor mergers and the accretion of numerous small companions play
important roles. Such lesser collision events are probably very common
in groups and clusters. As in solar system formation, ``core
accretion'' may be as important as monolithic collapse and
hierarchical buildup.

\subsection{Observations of Evolution}
\label{sec:22}

Let us return to observation. As in the case of induced SF discussed
above, there are two approaches - study of the statistical properties
of large samples, or study of individual objects in detail. In the
last decade there have been a great many surveys to provide data for
the first type of analysis (see the overview of
\citealt{iri04}). These include Hubble Space Telescope projects like
the Medium Deep Survey, the Hubble Deep Fields North and South, and
most recently GOODS (the Great Observatories Origins Deep Survey),
carried out in collaboration with the Chandra Observatory and the
Spitzer Space Telescope (see special issue of the Astrophysical
Journal Letters, Jan, 10, 2004), and GEMS (Galaxy Evolution from
Morphology and Spectral energy distributions, e.g., \citealt{bel05}).

The science and observing techniques of deep field survey and high
redshift studies in general are well beyond the scope of this
review. This is also not the place to consider the many different
classes of high redshift galaxies in any detail. These topics have
become the subject of wide interest and a burgeoning
literature. However, specific products of these studies, like the
merger rates, the cosmic star formation rate and mean morphological
statistics as a function of redshift, can provide information on the
history of galaxy collisions.

\subsubsection{Merger Rate versus Redshift}
\label{sec:23}

The differing predictions of the different models of galaxy formation,
and the interest in the role of major mergers/ULIRGs, have motivated a
continuing interest in the merger rate as a function of redshift. For
example, the CNOC cluster galaxy redshift project has reported quite
modest evolution in the merger rate at redshifts less than 1 (see
\citealt{pat02} and references therein). Specifically, Patton et
al. examined a sample of 4184 galaxies, found 88 galaxies in close
pairs, and derived a merger rate of $(1+z)^{2.3 \pm 0.7}$.

This result of low (and not rapidly changing) merger rate is confirmed
by several other recent studies, including the Caltech Faint Galaxy
Redshift Survey \citep{car00}. Moreover, \citet{lin04} report initial
results of the DEEP2 survey, in which they find a merger rate of
$(1+z)^m$ with exponent of $m = 0.51 \pm 0.28$ assuming mild
luminosity evolution, or $m = 1.6 \pm 0.29$ assuming no luminosity
evolution, since z = 1.2. They note that this implies only 9\% of
L$_*$ galaxies have undergone major mergers over this redshift
interval. Using deep infrared observations from the Subaru telescope
\citet{bun04} found that the fraction of close pairs (which usually
define the merger rate in these studies) increases ``modestly'' to
only about $7 \pm6$\% at z $\simeq$ 1. This is less than that found by
typical optical studies, and they note that the optical studies may be
``inflated'' by unrepresentative ``bright star-forming regions.''

Going in the other direction, \citet{lav04} find a very rapid increase
in the number of colliding ring galaxies with redshift. Head-on ring
galaxy collisions generally result in merger, so if the rings
represent a small randomly chosen fraction of all mergers, this would
imply a very rapidly evolving merger rate. On the other hand, if the
number of ring galaxies increases much more rapidly with redshift than
other types of merger, one wonders why?

For reference, we note that \citet{xu04} recently used data from the
2MASS near infrared survey to estimate the local merger rate; they
found the fraction of close major merger pairs to be $1.70 \pm
0.32$\%. For completeness, we note that at the time writing, mergers
rates based on the Sloan Digital Sky Survey or the 2dF survey have not
been published, though an initial atlas of SDSS merger pairs has
\citep{all04}. In the coming years it will be very interesting to see
statistically significant estimates of the merger rate extended to
well beyond z = 1.

\subsubsection{Cosmic Star Formation}
\label{sec:24}

Estimates of the mean SFR as a function of redshift are usually based
on color or emission line indicators (as opposed to the morphology
used to estimate merger rates in the pair surveys). In recent years
there have been surveys in a variety of wavebands. \citet{cra98}
carried out a novel radio continuum survey and found a local SFR of
about twice the optical H$\alpha$ value - 0.025 M$_\odot$ yr$^{-1}$
Mpc$^{-3}$. He found a value about 12 times greater at z $\simeq$ 1.
An analysis based on the 2dF survey also found an strong increase
($\propto (1+z)^b$, with $b < 5$) back to z $\simeq$ 1, and a more
moderate increase at redshifts of 1-5 \citep{bal02}. Analyses based on
SDSS data come to similar conclusions \citep{gla03, bri04}. The HST
STIS Parallel Survey found an SFR at z $\simeq$ 1 of $0.043 \pm 0.014$
M$_\odot$ yr$^{-1}$ Mpc$^{-3}$, based on [OII] emission \citep{tep03},
which is lower than most of the previous results.

The CADIS survey found SFR decreased by about a factor of 20 between
redshift 1.2 and the present, and the authors note the agreement of
their extinction corrected results with far infrared results
(\citealt{hip03}, also see the Herschel Telescope survey of
\citealt{gla04}). Results from the Gemini Deep Deep Survey indicate
that the SFR was about 6 times higher at z = 2 than at present
\citep{jun05}. One of the most dramatic changes in SFR was the factor
of 30 found in a GALEX (Galaxy Evolution Explorer satellite)
ultraviolet survey between z = 1 and the present
\citep{sch05}. Ultraviolet luminous galaxies may not very
representative of the cosmic SFR, but they could be related to
colliding galaxies.

The newest and deepest surveys indicate that SFR declines from a peak
at moderate redshifts to lower values at the highest
redshifts. \citet{bun04} identify and study 54 galaxies in the Hubble
Ultra Deep field and conclude that the SFR at z = 6 is about 6 times
less than at z $\simeq$ 3. \citet{hea04} come to similar
conclusions about the general history of SF on the basis of an
analysis of SDSS and other surveys. \citet{jun05} describe the
situation as a cosmic starburst at z $\simeq$ 2.

Very recently, survey results have revealed the phenomena of
``downsizing,'' wherein the most massive galaxies form first, and most
of the SF takes place in progressively smaller galaxies as time goes
on (e.g., \citealt{pog04}, \citealt{bou05}, \citealt{jun05},
\citealt{leb05}, \citealt{sha05}). \citet{bun05} argue (based in
part on GOODS data) that downsizing also proceeds from early to late
Hubble types, and that merging plays a key role. The implication is
that there is a mass dependence in the merger rate at any given epoch.

These new cosmic SFR results provide very interesting inputs to the
story of galaxy evolution. However, the relation of these results to
interactions and mergers remains to be clarified. Actually, the same
is true of the merger rate results, which are not sensitive to many
minor mergers or other interaction phenomena.

\subsubsection{Morphology versus Redshift}
\label{sec:25}

For an outsider the phenomenology of high redshift galaxies, which is
much constrained by detection techniques, is a daunting jungle of
jargon. Moreover, the relation between increasingly elaborate
simulations of structure formation and the observations is
complex. With rapid advances on both fronts, and increased efforts in
analysis and synthesis, we can expect much more clarity in the coming
decade (see review of \citealt{spi04} for a lucid current
discussion). For the present we focus on a few simple questions. Are
we directly observing galaxy evolution, i.e., the changing appearance
(build-up) of galaxies with redshift? Are collisions and mergers an
important part of this evolution?

\begin{figure}
\centering
%
\caption{\citet{con04b} sequence of ``low density objects''
at varying redshifts, illustrating the development of Hubble type
galaxies. (Courtesy C. Conselice)}
\label{fig:9}
\end{figure}

There is much new evidence in favor of an affirmative answer to both
questions (see commentary of \citealt{con04a}, with further details on
GOODS data in \citealt{con04b}). To say it a bit more emphatically,
these papers and those referenced within them suggest that we may be
beginning to acquire the observations that directly show the buildup
of typical Hubble sequence galaxies (see Figure 9).

There is much information to be found in the literature on the
properties of individual high redshift objects (individual galaxies
and clusters). We cannot review this literature here, and refer the
reader to the review of \citet{spi04}. Instead, let us return to the
subject of massive, or at least luminous, galaxies at high redshift,
and in particular, the interesting classes of extremely red galaxies
and submillimeter galaxies (or SCUBA galaxies, after the detector on
the James C. Maxwell Telescope). The latter are very infrared
luminous, high redshift objects (e.g., \citealt{con03},
\citealt{gen04}, \citealt{swi04}, \citealt{pop05}). Until recently,
only a few were known, but recent deep searches are beginning to
detect a substantial number (\citealt{gre04}, \citealt{sma04},
\citealt{wan04}). It appears that most of these objects are either
dust obscured quasars or high redshift LIRGs or ULIRGs, with perhaps
about 2/3 being the latter (\citealt{con03}, \citealt{ner03},
\citealt{sma04}, \citealt{swi04}). As with their local counterparts,
the LIRGs and ULIRGs are generally mergers or mergers-in-progress (see
\citealt{geo05}).

It appears that the submillimeter LIRGs are much more common than
present day LIRGS, and that they generate a substantial fraction of
the IR background (e.g., \citealt{gen04}, \citealt{wan04}). They have
typical redshifts of 2-3, and thus, coincide with the peak of the
cosmic SFR. They may be well represented among the most luminous
galaxies in that peak, but the evidence is preliminary. These results
are beginning to provide direct evidence that major mergers, if not
hierarchical buildup, were major contributors to galaxy evolution and
the cosmic star formation rate at these redshifts.

The submillimeter galaxies may be related to another high redshift
class, the Lyman Break Galaxies \citep{shu01}. However, few of the
latter are detected as the former (\citealt{cha00}, but note the
outstanding exception Westphal-MMD 11 discussed in
\citealt{cha02}). The deep ISO (Infrared Space Observatory) ELAIS
survey also found a number of ULIRGs at z $<$ 1 \citep{row04}. These
objects may bridge the gap between local ULIRGs and SCUBA
galaxies. Recent observations with the Spitzer Space Telescope show
that SCUBA galaxies are generally detectable at 24 microns, but with a
wide range of mid-infrared colors \citep{fra04}. Spitzer
observations also promise to delineate active nuclei from starburst
powered submillimeter galaxies \citep{ivi04}. All of this work
should contribute substantially to our understanding of the ``ULIRG
rate'' as a function of redshift (see the review of the cosmic
evolution of luminous infrared galaxies by \citealt{san04}).

Before submillimeter galaxies were discovered, observers were already
very interested in ``extremely red objects'' at high
redshift. Naively, one might expect to find more and more blue
galaxies at high redshift, and as described above, this is generally
the case. In this context, finding very red galaxies is surprising. On
the other hand, with a knowledge of dust-enshrouded starbursts in
ULIRGs, maybe this is not so surprising, but are EROs ULIRGs? Recent
studies suggest not, but rather many of them may be the already (at
typical redshifts of $>$ 1-2) old, red progenitors of present-day
early type galaxies (e.g., \citealt{fra03}, \citealt{for04},
\citealt{yan03}, \citealt{yan04a}, \citealt{yan04b} \citealt{bel05}).

Redness and age are relative terms. The typical age of the stellar
populations in these galaxies may be about 1 Gyr, which locally would
not be described as an old, red population. However, at the high
redshifts where these galaxies are found, the age of the universe when
the light was emitted was only a few Gyr or less.

Nonetheless, a fraction of the extremely red objects may be
ULIRGs. \citet{yan03} in an HST study of the morphology of a
sample at redshifts of about 1-2, estimate that about $17 \pm 4\%$ of the
objects are mergers or interactions. However, for the majority
dominated by older stellar populations we will have to seek merging
and interacting progenitors at still higher redshifts.

In conclusion, the above paragraphs describe the great advances that
have been made in recent years in studies of galaxy evolution at high
redshift. This work is impressive, but it is still hampered by
resolution, sensitivity and statistical limitations. There are hints
that mergers and interactions are important at all stages, but there
is a great deal more work to do before we understand the details.

\section{Archaeology}
\label{sec:26}

As discussed in several places above, models of particular classes of
collision have gotten quite sophisticated at this stage in the
development of the field. However, in the case of specific systems
this generalization is true only for systems that have experienced
only one close encounter, or where the time between encounters is so
long that the signatures of the first encounter have been largely
erased. The one notable exception to this caveat is the M51 system,
which may well be the result of two close encounters (see discussion
in \citealt{str99a}).

There are wave and tidal morphologies characteristic of cases with two
close passes separated by a time interval of order the mean internal
dynamical (e.g., rotational) time in the primary disk. \citet{str90}
demonstrated this explicitly in the case of colliding ring galaxies,
and it is quite clear in a number of merger models. It also seems
likely that a number of objects in the colliding galaxy atlases
require two close encounters to explain their morphology (e.g., the
M51 types).

Another kind of double encounter that has so far received only
exploratory attention is the case when a galaxy falls into a group and
has close encounters with more than one group member.

The study, and ultimately the classification, of double encounter
morphologies and merger remnant systematics is one that could advance
a long way in the next decade. There are no insurmountable technical
difficulties preventing advancement, though a great deal of numerical
effort will be required. Not only would a large number of simulations
have to be run, but they would have to be fully self-consistent
models. The ability to decipher two stages of development in colliding
systems would represent a substantial advance in galaxy archaeology.

\section{Coming Attractions}
\label{sec:27}

For the patient reader, it should be clear from the above that this
field has had a very exciting first fifty years or so. It must be
admitted that this is in large measure due to outside influences. Like
all other parts of astronomy, the study of galaxy collisions has
ridden the breaking waves of the vast technological advances in
detectors, satellite engineering, and computational resources. The
subject has received further boosts from the enormous interest in
parallel areas of study within the fields of galaxy evolution and star
formation, and has contributed back to those areas. Never again will
so many new, information-rich wavebands be opened. On the other hand,
wide scientific frontiers remain to be explored with the aid of
continuing increases in observational sensitivity, resolution,
computational power, and synergistic interactions with allied fields.

In the last few sections I have attempted to clarify where we stand on
the key questions posed at the end of Section 1. The first group of
questions concerned the role of galaxy collisions within the overall
picture of galaxy formation and evolution. Toomre's work in the 1970s
held out the possibility that collisions and interactions were a
dominant process, and that possibility has energized the field for
most of the time since. However, there have always been
counter-arguments. One of the strongest in the present era is the
modest increase in the merger rate with redshift found in deep
surveys. On the other hand, the relations between cosmic SFR or ULIRG
numbers and redshift tell a different story. Downsizing may be an
important part of the resolution between the different
stories. Presently, we only see a hazy outline of the full portrait of
the relation between galaxy morphology and redshift. Progress has been
rapid in these areas, and we can expect a great deal more in the next
decade or two. For optimistic theorists the answers are already
available (if not yet fully extracted) from large-scale numerical
models of structure formation.

The second group of key questions concerned the role of environment on
collision dynamics and evolutionary processes. Although the study of
galaxy collisions in groups and clusters has been around since Spitzer
and Baade's work, it is being reborn in the present era. There is
currently a great deal of interest in groups and clusters among
observers, with new tools to facilitate that work. The theory and
modeling side of this area is more complicated than that of binary
collisions and mergers because of the interaction between several
strong dynamical processes (e.g., ram pressure stripping,
group/cluster direct or tidal effects, etc.). Nonetheless, it is also
reasonable to expect significant advances in this area on decadal
timescales.

The third group of questions concerned the orchestration of SF and
nuclear activity by large-scale interaction dynamics. In the recent
past it seemed likely that progress in this area would be hindered by
the interplay of a number of complex dynamical processes. It now
appears that this view was overly pessimistic. Due to the universal
properties of turbulent interstellar gas, it now seems that wherever
you compress cool gas you will enhance SF (in quantifiable
ways). Thus, large-scale orchestration is mostly about gathering and
compressing gas; feedback effects are mostly about heating and
dispersing gas. There are more complexities than this, but the big
picture does not appear impossibly complicated. Work in the coming
decades should provide a much firmer foundation for this scenario, and
a much better understanding of the exceptional cases.

So the reviewer's crystal ball conveys a bright outlook for answers to
the first three groups of key questions. The glass gets more murky
when we ask the last couple of questions. The fourth group of
questions concerned secular effects and the fifth was about the
archaelogy of individual systems. Of course we can model long-term
processes on the computer with ever more precision. However, as
discussed in several contexts above, it is hard to compare to
observation either statistically or in individual cases.

That said, I would expect more progress from statistical studies, even
though that will require the relatively slow accumulation of good
datasets on numerous systems, for example acquiring large libraries of
faint tidal structures in numerous galaxies. That slow work is not
likely to be taken up by professional astronomers, but with the
increasing availability of moderate sized telescopes and sensitive CCD
detectors, it could become the realm of serious amateurs or robot
astronomers.

It seems possible that the majority of key questions discussed above
will be resolved within the next 50 years. However, new phenomena will
be discovered, and more detailed understandings will be
demanded. Recent, and possible near-future, examples support the
point. As an example, consider the exotic forms or products of SF,
like the ULX sources, and the possibility that some of these X-ray
sources are intermediate mass black holes formed in dense, young star
clusters. It will take a lot more observational work to explicate this
phenomenon, and probably new theoretical insights to explain it. We
could use several more Chandra observatories!

There are also still a few wavebands that remain largely unexplored,
including low-frequency radio waves, very high-energy gamma rays (new
more sensitive Cherenkov telescope arrays are presently coming on
line), and gravitational waves of many frequencies. Equally exciting
is the prospect that within a few decades our understanding of galaxy
disk hydrodynamics may advance to point that we understand both the
small scale, relatively short time, weather changes occurring in
isolated disks, and the longer term climate changes wrought by various
types of collisions and interactions. However, there is a great deal
of work to be done before goal is achieved.

\section{Acknowledgements}

I am very grateful to my research collaborators for teaching me much
about colliding galaxies and related topics. I want to thank Bev
Smith, in particular, for making a number of helpful suggestions on
this manuscript. I'd also like to acknowledge support from a NASA
Spitzer GO Cycle 1 grant.



%
%
%
%
%
%
%
%
%
%
\input{referenc}



\printindex
\end{document}

%% file: referenc.tex
%
%

%
%

%% file: cs.bbl
\begin{thebibliography}{99.}
%
%
%




\bibitem[Abadi, Moore \& Bower(1999)]{aba99} Abadi, M. G., Moore, B.,
\& Bower, R. G. (1999) Ram Pressure Stripping of Spiral Galaxies in
Clusters, MNRAS, \textbf{308}, 947

\bibitem[Allam et al.(2004)]{all04} Allam, S. S., et al. (2004)
Merging Galaxies in the Sloan Digital Sky Survey Early Data Release,
AJ, \textbf{127}, 1883

\bibitem[Alonso-Herrero et al.(2002)]{alo02} 
Alonso-Herrero, A., et al. (2002) Massive Star
Formation in Luminous Infrared Galaxies: Giant HII Regions and Their
Relation to Super-Star Clusters, AJ, \textbf{124}, 166

\bibitem[Appleton \& Struck-Marcell(1987a)]{app87a} 
Appleton, P. N., \& Struck-Marcell, C. (1987a) Star
Formation Rates in Ring Galaxies from IRAS Observations, ApJ,
\textbf{312}, 566

\bibitem[Appleton \& Struck-Marcell(1987b)]{app87b} 
Appleton, P. N., \& Struck-Marcell, C. (1987b)
Models of Ring Galaxies: II. Extended Starbursts, ApJ, \textbf{323},
480

\bibitem[Appleton \& Struck-Marcell(1996)]{app96a} 
Appleton, P. N., \& Struck-Marcell, C. (1996)
Collisional Ring Galaxies, Fun. Cos. Phys., \textbf{16}, 111

\bibitem[Appleton et al.(1996)]{app96b} 
Appleton, P. N, et al. (1996) Mapping Stellar
Evolution in the Wake of Density Waves in Ring Galaxies. In:
\textit{New Light on Galaxy Evolution, I.A.U. Symp. 171}, ed by
R. Bender \& R. L. Davies, (Kluwer, Dordrecht) p. 337

\bibitem[Arp(1966)]{arp66} 
Arp, H. (1966) Atlas of Peculiar Galaxies, ApJS,
\textbf{123}, 1

\bibitem[Arribas et al.(2004)]{arr04} 
Arribas, S., et al. (2004) Optical Imaging of Very
Luminous Infrared Galaxy Systems: Photometric Properties and Late
Evolution, AJ, \textbf{127}, 2522

\bibitem[Athanassoula(2004)]{ath04} 
Athanassoula, E. (2004) Dynamical Evolution of Barred
Galaxies, Amer. Astr. Soc. (DDA mtg.), \textbf{35}, \#0305 (astroph
0501196) 

\bibitem[Baade \& Minkowski(1954)]{baa54} 
Baade, W., \& Minkowski, R. (1954) Identification of
the Radio Sources in Cassiopeia A, Cygnus A, and Puppis A, ApJ,
\textbf{119}, 206

\bibitem[Baldry et al.(2002)]{bal02} 
Baldry, I. K., et al. (2002) The 2dF Galaxy Redshift
Survey: Constraints on Cosmic Star Formation History from the Cosmic
Spectrum, ApJ, \textbf{569}, 582

\bibitem[Barnes(2004)]{bar04}
Barnes, J. E. (2004) Shock-induced Star Formation in a
Model of the Mice, MNRAS, \textbf{350}, 798

\bibitem[Barnes \& Hernquist(1992)]{bar92} 
Barnes, J. E., \& Hernquist, L. (1992) Dynamics of
Interacting Galaxies, ARAA, \textbf{30}, 705

\bibitem[Barnes \& Hernquist(1996)]{bar96} 
Barnes, J. E., \& Hernquist, L. (1996) Transformations
of Galaxies. II. Gasdynamics in Merging Disk Galaxies, ApJ,
\textbf{471}, 115 

\bibitem[Barton, Geller, \& Kenyon(2000)]{bar00} 
Barton, E. J., Geller, M. J., \& Kenyon, S. J. (2000)
Tidally Triggered Star Formation in Close Pairs of Galaxies, ApJ,
\textbf{530}, 660 

\bibitem[Barton Gillespie, Geller, \& Kenyon(2003)]{bar03} 
Barton Gillespie, E. J., Geller, M. J., \& Kenyon,
S. J. (2003) Tidally Triggered Star Formation in Close Pairs of
Galaxies. II. Constraints on Burst Strengths and Ages, ApJ,
\textbf{582}, 668 

\bibitem[Bastian et al.(2005)]{bas05} 
Bastian, N., et al. (2005) The Star Cluster Population
of M51: II. Age Distribution and Relations Among the Derived
Parameters, A\&A, \textbf{431}, 905 

\bibitem[Begelman(2002)]{beg02} 
Begelman, M. C. (2002) Super-Eddington Fluxes from
Thin Accretion Disks?, ApJ, \textbf{568}, L97

\bibitem[Bekki(1997)]{bek97} 
Bekki, K. (1997) Formation of Polar Ring S0 Galaxies
in Dissipative Galaxy Mergers, ApJ, \textbf{490}, L37 

\bibitem[Bekki(2001)]{bek01} 
Bekki, K. (2001) Starbursts in Multiple Galaxy
Mergers, ApJ, \textbf{546}, 189 

\bibitem[Bekki et al.(2004)]{bek04} 
Bekki, K., et al. (2004) Formation of Star Clusters in
the Large Magellanic Cloud and Small Magellanic Cloud: I. Preliminary
Results on Cluster Formation From Colliding Gas Clouds, ApJ,
\textbf{602}, 730 

\bibitem[Bell et al.(2005)]{bel05} 
Bell, E. F., et al. (2005) Towards an Understanding of
the Rapid Decline of the Cosmic Star Formation Rate, ApJ,
\textbf{625}, 23 

\bibitem[Bergvall, Laurikainen, \& Aalto(2003)]{ber03} 
Bergvall, N., Laurikainen, E., \& Aalto, S. (2003)
Galaxy Interactions - Poor Starburst Triggers . III. A Study of a
Complete Sample of Interacting Galaxies, A\&A, \textbf{405}, 31 

\bibitem[Bianchi et al.(2005)]{bia05} 
Bianchi, L., et al. (2005) Recent Star Formation in Nearby Galaxies
from GALEX Imaging: M101 and M51, ApJ, \textbf{619}, L71

\bibitem[Bik et al.(2003)]{bik03} 
Bik, A., et al. (2003) Clusters in the Inner Spiral
Arms of M51: The Cluster IMF and the Formation History, A\&A,
\textbf{397}, 473 

\bibitem[Boselli et al.(2005)]{bos05} 
Boselli, A., et al. (2005) GALEX Ultraviolet
Observations of the Interacting Galaxy NGC 4438 in the Virgo Cluster,
ApJ, \textbf{623}, L13 

\bibitem[Bouche \& Lowenthal(2005)]{bou05} 
Bouche, N., \& Lowenthal, J. D. (2005) The Star
Formation Rate-Density Relationship at Redshift Three, ApJL,
\textbf{623}, L75 

\bibitem[Braine et al.(2004)]{bra04} 
Braine, J., et al. (2004) Colliding Molecular Clouds
in Head-on Galaxy Collisions, A\&A, \textbf{418}, 419 

\bibitem[Brinchman et al.(2004)]{bri04} 
Brinchman, J., et al. (2004) The Physical Properties
of Star Forming Galaxies in the Low Redshift Universe, MNRAS,
\textbf{351}, 1151 

\bibitem[Bundy, Ellis \& Conselice(2005)]{bun05} 
Bundy, K., Ellis, R. S., \& Conselice, C. J. (2005)
The Mass Assembly Histories of Galaxies of Various Morphologies in the
GOODS Fields, ApJ, \textbf{625}, 621 

\bibitem[Bundy et al.(2004)]{bun04} 
Bundy, K., et al. (2004) A Slow Merger History of
Field Galaxies since z $\simeq$ 1, ApJ, \textbf{601}, L123 

\bibitem[Bunker et al.(2004)]{bunk04} 
Bunker, A. J., et al. (2004) The Star Formation Rate
of the Universe at z $\simeq$ 6 from the Hubble Ultra Deep Field, MNRAS,
\textbf{355}, 374 

\bibitem[Burstein et al.(2004)]{bur04} 
Burstein, D., et al. (2004) Globular Cluster and
Galaxy Formation: M31, The Milky Way, and Implications for Globular
Cluster Systems of Spiral Galaxies, ApJ, \textbf{614}, 158

\bibitem[Bushouse(1987)]{bus87} 
Bushouse, H. A. (1987) Global Properties of
Interacting Disk-type Galaxies, ApJ, \textbf{320}, 49 

\bibitem[Carlberg et al.(2000)]{car00} 
Carlberg, R., et al. (2000) Caltech Faint Galaxy
Redshift Survey. XI. The Merger Rate to Redshift 1 from Kinematic
Pairs, ApJ, \textbf{532}, L1 

\bibitem[Chandrasekhar(1943)]{cha43} 
Chandrasekhar, S. (1943) Dynamical
Friction. I. General Considerations: the Coefficient of Dynamical
Friction, ApJ, \textbf{97}, 255 

\bibitem[Chapman et al.(2000)]{cha00} 
Chapman, S. C., et al. (2000) A Search for the
Submillimetre Counterparts to Lyman Break Galaxies, MNRAS,
\textbf{319}, 318 

\bibitem[Chapman et al.(2002)]{cha02} 
Chapman, S. C., et al. (2002) Westphal-MMD 11: An
Interacting, Submillimeter Luminous, Lyman Break Galaxy, ApJ,
\textbf{572}, L1 

\bibitem[Conselice(2004)]{con04a} 
Conselice, C. J. (2004) Unveiling the Formation of
Massive Galaxies, Science, \textbf{304}, 399 

\bibitem[Conselice, Chapman, \& Windhorst(2003)]{con03} 
Conselice, C. J., Chapman, S. C., \& Windhorst,
R. A. (2003) Evidence for a Major Merger Origin of High-Redshift
Submillimeter Galaxies, ApJ, \textbf{596}, L5 

\bibitem[Conselice et al.(2004)]{con04b} 
Conselice, C. J., et al. (2004) Observing the
Formation of the Hubble Sequence in the Great Observatories Origins
Deep Survey, ApJ, \textbf{600}, L139 

\bibitem[Corbett et al.(2003)]{cor03} 
Corbett, E. A., et al. (2003), COLA. II. Radio and
Spectroscopic Diagnostics of Nuclear Activity in Galaxies, ApJ,
\textbf{583}, 670 

\bibitem[Cox et al.(2005)]{cox05} 
Cox, T. J., et al. (2005) The Effects of Feedback in
Simulations of Disk-galaxy Major Mergers, MNRAS, submitted (astro-ph
0503201) 

\bibitem[Cram(1998)]{cra98} 
Cram, L. E. (1998) The Global Star Formation Rate from
the 1.4 GHz Luminosity Funcition, ApJ, \textbf{506}, 85 

\bibitem[De Grijs(2001)]{deg01} 
De Grijs, R. (2001) Star Formation Timescales in M82,
Astr. \& Geophys., \textbf{42}, 12 

\bibitem[De Grijs, O'Connell, \& Gallagher(2002)]{deg02} 
De Grijs, R., O'Connell, R. W., \& Gallagher III,
J. S. (2002) Tidally-induced Super Star Clusters in M82, in
\textit{Extragalactic Star Clusters}, I.A.U. Symp. \textbf{207},
eds. D. Geisler, E. K. Grebel, and D. Minniti, (ASP, San Francisco)
p. 477

\bibitem[Duc et al.(2000)]{duc00} 
Duc, P.-A., et al. (2000) Formation of a Tidal Dwarf
Galaxy in the Interacting System Arp 245 (NGC 2992(93), AJ,
\textbf{120}, 1238 

\bibitem[Duc et al.(1997)]{duc97} 
Duc, P.-A., et al. (1997) Gas Segregation in the
Interacting System Arp 105, A\&A, \textbf{326}, 537 

\bibitem[Duc, Bournaud, \& Masset(2004)]{duc04a} 
Duc, P.-A., Bournaud, F., \& Masset, F. (2004) A
Top-down Scenario for the Formation of Massive Tidal Dwarf Galaxies,
A\&A, \textbf{427}, 803 

\bibitem[Duc, Braine, \& Brinks(2004)]{duc04b} 
Duc, P.-A., Braine, J., \& Brinks, E. (2004)
\textit{Recycling Intergalactic and Interstellar Matter},
I.A.U. Symp. \textbf{217}, (A.S.P., San Francisco) 

\bibitem[Ellingson et al.(2001)]{ell01} 
Ellingson, et al.(2001) The Evolution of Population
Gradients in Galaxy Clusters: The Butcher-Oemler Effect and Cluster
Infall, ApJ, \textbf{547}, 609 

\bibitem[Elmegreen et al.(2000)]{elm00} 
Elmegreen, B. G., et al. (2000) Hubble Space Telescope
Observations of the Interacting Galaxies NGC 2207 and IC 2163, AJ,
\textbf{120}, 630 

\bibitem[Elmegreen et al.(1991)]{elm91} 
Elmegreen, D. M., et al. (1991) Properties and
Simulations of Interacting Spiral Galaxies with Transient "Ocular"
Shapes, A\&A, \textbf{244}, 52 

\bibitem[Fabbiano et al.(2004)]{fab04} 
Fabbiano, G., et al. (2004) X-raying Chemical Evolution and Galaxy
Formation in the Antennae, ApJ, \textbf{605}, L21

\bibitem[Feldmeier et al.(2002)]{fel02} 
Feldmeier, J. J., et al. (2002) Deep CCD Photometry of
Galaxy Clusters. I. Methods and Initial Studies of Intracluster
Starlight, ApJ, \textbf{575}, 779

\bibitem[Fiorito \& Titarchuk(2004)]{fio04} 
Fiorito, R., \& Titarchuk, L. (2004) Is M82 X-1 Really
an Intermediate-Mass Black Hole? X-ray Spectral and Timing Evidence,
ApJ, \textbf{614}, L113 

\bibitem[F{\"o}rster Schreiber et al.(2004)]{for04} 
F{\"o}rster Schreiber, N. M., et al. (2004) A
Substantial Population of Red Galaxies at z>2: Modeling of the
Spectral Energy Distributions of an Extended Sample, ApJ,
\textbf{616}, 40 

\bibitem[Franx et al.(2003)]{fra03} 
Franx, M., et al. (2003) A Significant Population of
Red, Near-infrared-selected High-redshift Galaxies, ApJ, \textbf{587},
L79 

\bibitem[Frayer et al.(2004)]{fra04}
 Frayer, D. T., et al. (2004) Infrared Properties of Radio-selected
Submillimeter Galaxies in the Spitzer First Look Survey Verification
Field, ApJS, \textbf{154}, 137

\bibitem[Gao \& Solomon(1999)]{gao99} 
Gao, Y., \& Solomon, P. M. (1999) Molecular Gas Depletion and
Starbursts in Luminous Infrared Galaxy Mergers, ApJ,
\textbf{512}, L99 

\bibitem[Gao \& Solomon(2004)]{gao04} 
Gao, Y., \& Solomon, P. M. (2004) The Star Formation
Rate and Dense Molecular Gas in Galaxies, ApJ, \textbf{606}, 271 

\bibitem[Gao et al.(2003)]{gao03} 
Gao, Y., et al. (2003) Nonnuclear Hyper/Ultraluminous
X-ray Sources in the Starbursting Cartwheel Ring Galaxy, ApJ,
\textbf{596}, L171 

\bibitem[Genzel et al.(2004)]{gen04} 
Genzel, R., et al. (2004) Submillimeter Galaxies as
Tracers of Mass Assembly at Large M, astro-ph 0403183 

\bibitem[Georgakakis et al.(2005)]{geo05} 
Georgakakis, A., et al. (2005) Dust in a Merging Galaxy Sequence: the
SCUBA View, in \textit{The Spectral Energy Distribution of Gas-rich
Galaxies: Confronting Models with Data}, eds. C. C. Popescu and
R. J. Tuffs, (A.I.P. Conf. Series \textbf{761}, New York), 441

\bibitem[Glazebrook et al.(2003)]{gla03} 
Glazebrook, K., et al. (2003) The Sloan Digital Sky
Survey: The Cosmic Spectrum and Star Formation History, ApJ,
\textbf{587}, 55 

\bibitem[Glazebrook et al.(2004)]{gla04} 
Glazebrook, K., et al. (2004) Cosmic Star Formation
History to z=1 from a Narrow Emission Line-selected Tunable-filter
Survey, AJ, \textbf{128}, 2652 

\bibitem[Gnedin(2003a)]{gne03a} 
Gnedin, O. Y. (2003a) Tidal Effects in Clusters of
Galaxies, ApJ, \textbf{582}, 141  

\bibitem[Gnedin(2003b)]{gne03b} 
Gnedin, O. Y. (2003b) Dynamical Evolution of Galaxies
in Clusters, ApJ, \textbf{589}, 752 

\bibitem[Goto(2005)]{got05} 
Goto, T. (2005) 266 E+A Galaxies Selected from the
Sloan Digital Sky Survey Data Release 2: The Origin of E+A Galaxies,
MNRAS, \textbf{357}, 937 

\bibitem[Greve et al.(2004)]{gre04} 
Greve, T. R., et al. (2004) A 1200-mu MAMBO Survey of
ELAISN2 and the Lockman Hole - I. Maps, Sources and Number Counts,
MNRAS, \textbf{354}, 779 

\bibitem[Guti{\'e}rrez \& L{\'o}pez(2005)]{gut05} 
G{\'u}tierrez, C. M., \& L{\'o}pez-Corredoira, M. (2005) The
Nature of Ultra Luminous X-ray Sources, ApJ Letters, \textbf{622}, L89 

\bibitem[Harris et al.(2004)]{har04} 
Harris, J., et al. (2004) The Recent Cluster Formation
Histories of NGC 5253 and NGC 3077: Environmental Impact on Star
Formation, ApJ, \textbf{603}, 503 

\bibitem[Hashimoto \& Oemler(2000)]{has00} 
Hashimoto, Y., \& Oemler Jr., A. (2000) The Effect of
Environment on Galaxy Interactions, ApJ, \textbf{530}, 652 

\bibitem[Heavens et al.(2004)]{hea04} 
Heavens, A., et al. (2004) The Complete Star Formation
History of the Universe, Nature, \textbf{428}, 625 

\bibitem[Hernquist \& Quinn(1988)]{her88} 
Hernquist, L., \& Quinn, P. J. (1988) Formation of
Shell Galaxies: I. Spherical Potentials, ApJ, \textbf{331}, 682 

\bibitem[Hibbard \& Barnes(2004)]{hib04} 
Hibbard, J. E., \& Barnes, J. E. (2004) The Dynamical
Masses of Tidal Dwarf Galaxies, in \textit{Recycling Intergalactic and
Interstellar Matter}, I.A.U. Symp. \textbf{217}, eds. P.-A. Duc,
J. Braine, and E. Brinks, ASP, p. 510 

\bibitem[Hibbard, Rupen, \& van Gorkom(2001)]{hib01} 
Hibbard, J. E., Rupen, M. \& van Gorkom, J. H.,
eds. (2001) \textit{Gas and Galaxy Evolution}, ASP Conf. Proceedings
\textbf{240}, Astronomical Society of the Pacific 

\bibitem[Hibbard, Vacca, \& Yun(2000)]{hib00} 
Hibbard, J. E., Vacca, W. D., \& Yun, M. S. (2000) The
Neutral Hydrogen Distribution in Merging Galaxies: Differences Between
Stellar and Gaseous Tidal Morphologies, AJ, \textbf{119}, 1130 

\bibitem[Hippelein et al.(2003)]{hip03} 
Hippelein, H., et al. (2003) Star Forming Rates
between z = 0.25 and z = 1.2 from the CADIS Emission Line Survey,
A\&A, \textbf{402}, 65 

\bibitem[Hopman et al.(2004)]{hop04}
Hopman, C., et al. (2004) Ultraluminous X-ray Sources as
Intermediate-Mass Black Holes Fed by Tidally Captured Stars, ApJ,
\textbf{604}, L101

\bibitem[Irion(2004)]{iri04} 
Irion, R. (2004) Surveys Scour Cosmic Deep, Science
(News Note), \textbf{303}, 1750 

\bibitem[Ivison et al.(2004)]{ivi04} 
Ivison, R. J., et al. (2004) Spitzer Observations of
MAMBO Galaxies: Weeding Out Active Nuclei from Starbursting
Protoellipticals, ApJS, \textbf{154}, 124 

\bibitem[Javiel, Santiago, \& Kerber(2005)]{jav05} 
Javiel, S. C., Santiago, B. X., \& Kerber,
L. O. (2005) Constraints on the Star Formation History of the Large
Magellanic Cloud, A\&A, \textbf{431}, 73

\bibitem[Jog \& Solomon(1992)]{jog92} 
Jog, C. J., and Solomon, P. M. (1992) A Triggering
Mechanism for Enhanced Star Formation in Colliding Galaxies, ApJ,
\textbf{387}, 152 

\bibitem[Juneau et al.(2005)]{jun05} 
Juneau, S., et al. (2005) Cosmic Star Formation
History and Its Dependence on Galaxy Stellar Mass, ApJ, \textbf{619},
L135 

\bibitem[Keel(1993)]{kee93} 
Keel, W. C. (1993), Kinematic Regulation of Star
Formation, AJ, \textbf{106}, 1771 

\bibitem[Keel \& Borne(2003)]{kee03} 
Keel, W. C., \& Borne, K. D. (2003) Massive Star Clusters in Ongoing
Galaxy Interactions: Clues to Cluster Formation, AJ, \textbf{126},
1257

\bibitem[Keel et al.(1985)]{kee85} 
Keel, W. C., et al. (1985), The Effects of
Interactions on Spiral Galaxies. I. Nuclear Activity and Star
Formation, AJ, \textbf{90}, 708 

\bibitem[Kenney \& Yale(2002)]{ken02} 
Kenney, J. D. P., \& Yale, E. E. (2002) Hubble Space
Telescope Imaging of Bipolar Nuclear Shells in the Disturbed Virgo
Cluster Galaxy NGC 4438, ApJ, \textbf{567}, 865 

\bibitem[Kennicutt(1989)]{ken89} 
Kennicutt Jr., R. C. (1989) The Star Formation Law in
Galactic Disks, ApJ, \textbf{344}, 685 

\bibitem[Kennicutt(1998)]{ken98} 
Kennicutt Jr., R. C. (1998) The Global Schmidt Law in
Star-forming Galaxies, ApJ, \textbf{498}, 541 

\bibitem[Kennicutt et al.(1987)]{ken87} 
Kennicutt Jr., R. C., et al. (1987) The Effects of
Interactions on Spiral Galaxies. II. Disk Star-Formation Rates, AJ,
\textbf{93}, 1011 

\bibitem[King(2004)]{kin04} 
King, A. R. (2004) Ultraluminous X-ray Sources and
Star Formation, MNRAS, \textbf{347}, L18 

\bibitem[King \& Dehnen(2005)]{kin05} 
King, A. R., \& Dehnen, W. (2005) Hierarchical
Merging, Ultraluminous and Hyperluminous X-ray Sources, MNRAS,
\textbf{357}, 275 

\bibitem[Klimanov \& Reshetnikov(2001)]{kli01} 
Klimanov, S. A., \& Reshetnikov, V. P. (2001) A
Statistical Study of M51-type Galaxies, A\&A, \textbf{378}, 428 

\bibitem[Klypin et al.(1999)]{kly99} 
Klypin, A., et al. (1999) Where are the Missing
Galactic Satellites?, ApJ, \textbf{522}, 82

\bibitem[Knierman et al.(2003)]{kni03} 
Knierman, K. A., et al. (2003) From Globular Clusters
to Tidal Dwarfs: Structure Formation in the Tidal Tails of Merging
Galaxies, AJ, \textbf{126}, 1227 

\bibitem[Krolik(2004)]{kro04} 
Krolik, J. H. (2004) Are Ultraluminous X-ray Sources
Intermediate-Mass Black Holes Accreting from Molecular Clouds?, ApJ,
\textbf{615}, 383 

\bibitem[Krumholz \& McKee(2005)]{kru05} 
Krumholz, M. R., \& McKee, C. F., (2005) A General
Theory of Turbulence-Regulated Star Formation, From Spirals to ULIRGs,
ApJ, \textbf{630}, 250

\bibitem[Lambas et al.(2003)]{lam03} 
Lambas, D. G., et al. (2003) Galaxy Pairs in the 2dF
Survey - I. Effects of Interactions on Star Formation in the Field,
MNRAS, \textbf{346}, 1189 

\bibitem[Lan{\c{c}}on et al.(2001)]{lan01} 
Lan{\c{c}}on, A., et al. (2001) Multiwavelength Study
of the Starburst Galaxy NGC 7714. II. The Balance Between Young,
Intermediate-Age, and Old Stars, ApJ, \textbf{552}, 150 

\bibitem[Larson \& Tinsley(1978)]{lar78} 
Larson, R. B., \& Tinsley, B. M. (1978) Star Formation
Rates in Normal and Peculiar Galaxies, ApJ, \textbf{219}, 46 

\bibitem[Larson, Tinsley, \& Caldwell(1980)]{lar80} 
Larson, R. B., Tinsley, B. M., \& Caldwell,
C. N. (1980) The Evolution of Disk Galaxies and the Origin of SO
Galaxies, ApJ, \textbf{237}, 692 

\bibitem[Lavery et al.(2004)]{lav04} 
Lavery, R. J., et al. (2004) Probing the Evolution of
the Galaxy Interaction/Merger Rate Using Collisional Ring Galaxies,
ApJ, \textbf{612}, 679 

\bibitem[Le Borgne et al.(2005)]{leb05} 
Le Borgne, D., et al. (2005) Gemini Deep Deep Survey
VI: Massive Post-starburst Galaxies at z=1, ApJ, submitted (astro-ph
0503401) 

\bibitem[Lewis, Buote, \& Stocke(2003)]{lew03} 
Lewis, A. D., Buote, D. A., \& Stocke, J. T., (2003)
Chandra Observations of A2029: The Dark Matter Profile Down to below
0.01 r$_{VIR}$ in an Unusually Relaxed Cluster, ApJ, \textbf{586}, 135 

\bibitem[Lin et al.(2004)]{lin04} 
Lin, L., et al. (2004) The DEEP2 Galaxy Redshift
Survey: Evolution of Close Galaxy Pairs and Major-Merger Rates up to z
$\sim$ 1.2, ApJ, \textbf{617}, L9 

\bibitem[Liu, Bregman, \& Irwin(2005)]{liu05} 
Liu, J.-F., Bregman, J. N., \& Irwin, J. (2005)
Ultra-luminous X-ray Sources in Nearby Galaxies from ROSAT HRI
Observations, II. Statistical Properties, ApJS, \textbf{157}, 59

\bibitem[Lonsdale Persson(1986)]{lon86} 
Lonsdale Persson, C. J., ed., (1986), \textit{Star
Formation in Galaxies}, NASA Conf. Pub. 2466 

\bibitem[Mao et al.(2000)]{mao00} 
Mao, R. O., et al. (2000) Dense Gas in Nearby
Galaxies. XIII. CO Submillimeter Line Emission from the Starburst
Galaxy M 82, A\&A, \textbf{358}, 433 

\bibitem[Matsumoto et al.(2004)]{mat04} 
Matsumoto, H., et al. (2004) Peculiar Characteristics
of the Hyper-luminous X-ray Source M82 X-1, Prog. Th. Phys. Suppl.,
\textbf{155}, 379 

\bibitem[Melo(2005)]{mel05} 
Melo, V. P., (2005) Young Super Star Clusters in the
Starburst of M82: The Catalog, AJ, \textbf{619}, 270 

\bibitem[Mihos(2004)]{mih04} 
Mihos, J. C. (2004) Interactions and Mergers of
Cluster Galaxies, in  \textit{Clusters of Galaxies: Probes of
Cosmological Structure and Galaxy Evolution},  eds. J. S. Mulchaey,
A. Dressler, \& A. Oemler, (Cambridge Univ. Press, Cambridge) p. 277 

\bibitem[Mihos, Bothun, \& Richstone(1993)]{mih93} 
Mihos, J. C., Bothun, G. D., \& Richstone,
D. O. (1993) Modeling the Spatial Distribution of Star Formation in
Interacting Disk Galaxies, ApJ, \textbf{418}, 82 

\bibitem[Mihos \& Hernquist(1994)]{mih94} 
Mihos, J. C., \& Hernquist, L. (1994) Star-forming
Galaxy Models: Blending Star Formation into TREESPH, ApJ,
\textbf{437}, 611 

\bibitem[Mihos \& Hernquist(1996)]{mih96} 
Mihos, J. C., \& Hernquist, L. (1996) Gasdynamics and
Starbursts in Major Mergers, ApJ, \textbf{464}, 641 

\bibitem[Miller(2005)]{mil05} 
Miller, J. M. (2005) Present Evidence for Intermediate
Mass Black Holes in ULXs and Future Prospects, in \textit{From X-ray
Binaries to Quasars: Black Hole Accretion on All Mass Scales},
eds. T. J. Maccarone, R. P. Fender, \& L. C. Ho, (Kluwer, Dordrecht),
in press (astro-ph 0412526) 

\bibitem[Miller, Fabian, \& Miller(2004)]{mil04a} 
Miller, J. M., Fabian, A. C., \& Miller, M. C. (2004)
A Comparison of Intermediate-Mass Black Hole Candidate Ultraluminous
X-ray Sources and Stellar-Mass Black Holes, ApJ, \textbf{614}, L117 

\bibitem[Miller et al.(2004)]{mil04b} 
Miller, J. M., et al. (2004) XMM-Newton Spectroscopy
in the Antennae Galaxies (NGC 4038/4039), ApJ, \textbf{609}, 728 

\bibitem[Moore et al.(1996)]{moo96} 
Moore, B., et al. (1996) Galaxy Harrassment and the
Evolution of Clusters of Galaxies, Nature, \textbf{379}, 613 

\bibitem[Moore et al.(1999)]{moo99} 
Moore, B., et al. (1999) On the Survival and
Destruction of Spiral Galaxies in Clusters, MNRAS, \textbf{304}, 465 

\bibitem[Mushotsky(2004a)]{mus04a} 
Mushotsky, R. (2004a) Ultra-luminous Sources in
Nearby Galaxies, Prog. Th. Phys. Suppl., \textbf{155}, 27 

\bibitem[Mushotsky(2004b)]{mus04b} 
Mushotsky, R.. (2004b) Clusters of Galaxies: an X-ray
Perspective, in \textit{Clusters of Galaxies: Probes of Cosmological
Structure and Galaxy Evolution},  eds. J. S. Mulchaey, A. Dressler, \&
A. Oemler, (Cambridge Univ. Press, Cambridge) p. 134 

\bibitem[Nagamine et al.(2005)]{nag05} 
Nagamine, K., et al. (2005) Massive Galaxies in
Cosmological Simulations: Ultraviolet-selected Sample at Redshift z=2,
ApJ, \textbf{618}, 23 

\bibitem[Navarro, Frenk, \& White(1997)]{nav97} 
Navarro, J. F., Frenk, C. S., \& White,
S. D. M. (1997) A Universal Density Profile from Hierarchical
Clustering, ApJ, \textbf{490}, 493 

\bibitem[Neri et al.(2003)]{ner03} 
Neri, R., et al. (2003) Interferometric Observations
of Powerful CO Emission from Three Submillimeter Galaxies at z = 2.39,
2.51, and 3.35, ApJ, \textbf{597}, L113 

\bibitem[Nikolic, Cullen, \& Alexander(2004)]{nik04} 
Nikolic, B., Cullen, H., \& Alexander, P. (2004) Star
Formation in Close Pairs Selected from the Sloan Digital Sky Survey,
MNRAS, \textbf{355}, 874 

\bibitem[Noguchi(1987)]{nog87} 
Noguchi, M. (1987) Close Encounter Between Galaxies:
II. Tidal Deformation of a Disc Galaxy Stabilized by a Massive Halo,
MNRAS, \textbf{228}, 635 

\bibitem[O'Connell(2004)]{oco04} 
O'Connell, R. W. (2004) Ten Years of Super Star
Cluster Research, in \textit{The Formation and Evolution of Massive
Young Clusters}, A.S.P. Conf. \textbf{322},
eds. H. J. G. L. M. Lamers, L. J. Smith, \& A. Nota, (A.S.P., San
Francisco) p. 551 

\bibitem[Patton et al.(2002)]{pat02} 
Patton, D. R., et al. (2002) Dynamically Close Galaxy
Pairs and Merger Rate Evolution in the CNOC2 Redshift Survey, ApJ,
\textbf{565}, 208 

\bibitem[Pimbblet(2003)]{pim03} 
Pimbblet, K. (2003) At the Vigintennial of the
Butcher-Oemler Effect, PASA, \textbf{20}, 294 

\bibitem[Poggianti et al.(2004)]{pog04} 
Poggianti, B. M., et al. (2004) A Comparison of the
Galaxy Populations in the Coma and Distant Clusters: The Evolution of
the k+a Galaxies and the Role of the Intracluster Medium, ApJ,
\textbf{601}, 197 

\bibitem[Pope et al.(2005)]{pop05} 
Pope, A., et al. (2005) The Hubble Deep Field North
SCUBA Super-map III - Optical and Near-infrared Properties of
Submillimetre Galaxies, MNRAS, \textbf{358}, 149 

\bibitem[Portegies Zwart, Dewi, \& Maccarone(2004)]{por04}
Portegies Zwart, S., Dewi, J., \& Maccarone, T. (2004)
Intermediate Mass Black Holes in Accreting Binaries: Formation,
Evolution and Observational Appearance, MNRAS, \textbf{355}, 413

\bibitem[Roediger \& Hensler(2005)]{roe05} 
Roediger, E., \& Hensler, G. (2005) Ram Pressure
Stripping of Disk Galaxies A\&A, in press (astro-ph 0412518) 

\bibitem[Rowan-Robinson et al.(2004)]{row04} 
Rowan-Robinson, M., et al. (2004) The European
Large-area ISO Survey (ELAIS): the Final Band-merged Catalogue, MNRAS,
\textbf{351}, 1290 

\bibitem[Sandage(1961)]{san61} 
Sandage, A. (1961) \textit{The Hubble Atlas of
Galaxies}, (Carnegie Inst. of Washington, Washington) (Pub. 618) 

\bibitem[Sanders(2004)]{san04} 
Sanders, D. B. (2004) The Cosmic Evolution of Luminous
Infrared Galaxies: from IRAS to ISO, SCUBA, and SIRTF, AdSpR,
\textbf{34}, 535 

\bibitem[Sanders \& Mirabel(1996)]{san96} 
Sanders, D. B., \& Mirabel, I. F. (1996) Luminous
Infrared Galaxies, ARAA, \textbf{34}, 749 

\bibitem[Scannapieco, Weisheit, \& Harlow(2004)]{sca04} 
Scannapieco, E., Weisheit, J., \& Harlow, F. (2004)
Triggering the Formation of Halo Globular Clusters with Galaxy
Outflows, ApJ, \textbf{615}, 29 

\bibitem[Schiminovich et al.(2005)]{sch05} 
Schiminovich, D., et al. (2005) The GALEX-VVDS
Measurement of the Evolution of the Far Ultraviolet Luminosity Density
and the Cosmic Star Formation Rate, ApJ, \textbf{619}, L47 

\bibitem[Schulz \& Struck(2001)]{sch01} 
Schulz, S., \& Struck, C. (2001) Multi Stage
Three-dimensional Sweeping and Annealing of Disc Galaxies in Clusters,
MNRAS, \textbf{328}, 185 

\bibitem[Schweizer(1983)]{schw83} 
Schweizer, F. (1983) Observational Evidence for
Mergers, in \textit{Internal Kinematics and Dynamics of Galaxies},
ed. E. Athanassoula, (Reidel, Dordrecht) p. 319 

\bibitem[Schweizer(1998)]{schw98} 
Schweizer, F. (1998) Observational Evidence for
Interactions and Mergers, in \textit{Galaxies: Interactions and
Induced Star Formation}, Saas Fee Advanced Course \textbf{26},
eds. D. Friedli, D. Martinet, \& D. Pfenniger, (Springer, Berlin)
p. 105 

\bibitem[Schweizer(2005)]{schw05} 
Schweizer, F. (2005) in \textit{Starbursts: From 30
Doradus to Lyman Break Galaxies}, eds. R. de Grijs \& R. M. Gonzalez
Delgado, (Kluwer, Dordrecht) p. 143 (astro-ph 0502111) 

\bibitem[Schombert, Wallin, \& Struck-Marcell(1990)]{scho90} 
Schombert, J. M., Wallin, J. F., \& Struck-Marcell,
C. (1990) A Multicolor Photometric Study of the Tidal Features in
Interacting Galaxies, AJ, \textbf{99}, 497 

\bibitem[Shapley et al.(2005)]{sha05} 
Shapley, A. E., et al. (2005) UV to Mid-IR
Observations of Star-forming Galaxies at z~2: Stellar Masses and
Stellar Populations, ApJ, \textbf{626}, 698

\bibitem[Shu, Mao, \& Mo(2001)]{shu01} 
Shu, C., Mao, S., \& Mo, H. J. (2001) The Host Haloes
o f Lyman-break Galaxies and Submillimetre Sources, MNRAS,
\textbf{327}, 895 

\bibitem[Smail et al.(2004)]{sma04} 
Smail, I., et al. (2004) The Rest-Frame Optical
Properties of SCUBA Galaxies, ApJ, \textbf{616}, 71 

\bibitem[Smith, Struck, \& Nowak(2005)]{smi05} 
Smith, B. J., Struck, C., \& Nowak, M. A. (2005)
Chandra X-ray Imaging of the Interacting Starburst Galaxy NGC
7714/7715: Tidal Ultra-luminous X-ray Sources, Emergent Wind, and
Resolved HII Regions, AJ, \textbf{129}, 1350 

\bibitem[Sofue \& Rubin(2001)]{sof01} 
Sofue, Y., \& Rubin, V. (2001) Rotation Curves of
Spiral Galaxies, ARAA, \textbf{39}, 137 

\bibitem[Soifer, Houck, \& Neugebauer(1987)]{soi87} 
Soifer, B. T., Houck, J. R., \& Neugebauer, G. (1987)
The IRAS View of the Extragalactic Sky, ARAA, \textbf{25}, 187 

\bibitem[Sparke(2002)]{spa02} 
Sparke, L. (2002) Off-plane Gas and Galaxy Disks, in
\textit{Disks of Galaxies: Kinematics, Dynamics and Perturbations},
ASP Conf. Series \textbf{275}, eds. E. Athanassoula, A. Bosma, \&
R. Mujica, (ASP, San Francisco) p. 367 

\bibitem[Spinrad(2004)]{spi04} 
Spinrad, H. (2004) The Most Distant Galaxies, in
\textit{Astrophysics Update: Topical and Timely Reviews in
Astrophysics}, ed. J. W. Mason, (Springer Praxis, Berlin) p. 155 

\bibitem[Spitzer \& Baade(1951)]{spi51}
Spitzer, L., Jr., \& Baade, W. (1951) Stellar
Populations and Collisions of Galaxies, ApJ, \textbf{113}, 413

\bibitem[Springel \& Hernquist(2005)]{spr05} 
Springel, V., \& Hernquist, L. (2005) Formation of a
Spiral Galaxy in a Major Merger, ApJ, \textbf{622}, L9 

\bibitem[Struck(1997)]{str97} 
Struck, C. (1997) Simulations of Collisions Between
Two Gas-Rich Galaxy Disks with Heating and Cooling, ApJS,
\textbf{113}, 269 

\bibitem[Struck(1999)]{str99a} 
Struck, C. (1999) Galaxy Collisions, Phys Rep,
\textbf{321}, 1 

\bibitem[Struck(2004)]{str04} 
Struck, C. (2004) Case Studies of Mass Transfer and
Star Formation in Galaxy Collisions, in \textit{Recycling
Intergalactic and Interstellar Matter}, I.A.U. Symp. \textbf{\#217},
eds. P.-A. Duc, J. Braine, \& E. Brinks, (ASP, San Francisco) p. 400 

\bibitem[Struck(2005)]{str05} 
Struck, C. (2005) The Recurrent Nature of Central
Starbursts, in \textit{Starbursts: From 30 Doradus to Lyman Break
Galaxies}, eds. R. de Grijs \& R. M. Gonzalez Delgado, (Kluwer,
Dordrecht) p. 163 

\bibitem[Struck \& Smith(2003)]{str03} 
Struck, C. \& Smith, B. J. (2003) Models of the
Morphology, Kinematics and Star Formation History of the Prototypical
Collisional Starburst System: NGC 7714/7715 = Arp 284, ApJ,
\textbf{589}, 157 

\bibitem[Struck \& Smith(1999)]{str99b} 
Struck, C., \& Smith, D. C. (1999) Simple Models for
Turbulent Self-Regulation in Galaxy Disks, ApJ, \textbf{527}, 673 

\bibitem[Struck-Marcell \& Lotan(1990)]{str90} 
Struck-Marcell, C., \& Lotan, P. (1990) The Varieties
of Symmetric Stellar Rings and Radial Caustics in Galaxy Disks, ApJ,
\textbf{358}, 99. 

\bibitem[Struck-Marcell \& Tinsley(1978)]{str78} 
Struck-Marcell, C., \& Tinsley, B. M. (1978) Star
Formation Rates and Infrared Radiation, ApJ, \textbf{221}, 562 

\bibitem[Sulentic, Keel, \& Telesco(1989)]{sul89} 
Sulentic, J. W., Keel, W. C., \& Telesco, C. M., eds.,
(1989) \textit{Paired and Interacting Galaxies:}
I.A.U. Colloq. No. \textbf{124}, NASA Conf. Pub. 3098 

\bibitem[Swartz et al.(2004)]{swa04} 
Swartz, D. A., et al. (2004) The Ultra-luminous X-ray
Source Population from the Chandra Archive of Galaxies, ApJS,
\textbf{154}, 519 

\bibitem[Swinbank et al.(2004)]{swi04} 
Swinbank, A. M., et al.(2004) The Rest-frame Optical
Spectra of SCUBA Galaxies, ApJ, \textbf{617}, 64 

\bibitem[Teplitz et al.(2003)]{tep03} 
Teplitz, H. I., et al. (2003) Emission-line Galaxies
in the STIS Parallel Survey. II. Star Formation Density, ApJ,
\textbf{589}, 704 

\bibitem[Toomre(1977)]{too77} 
Toomre, A. (1977) Mergers and Some Consequences, in
\textit{The Evolution of Galaxies and Stellar Populations},
eds. B. M. Tinsley \& R. B. Larson, (Yale University Observatory, New
Haven) p 401 

\bibitem[Toomre \& Toomre(1972)]{too72} 
Toomre, A., \& Toomre, J. (1972) Galactic Bridges and
Tails, ApJ, \textbf{178}, 623 

\bibitem[Tremaine \& Weinberg(1984)]{tre84} 
Tremaine, S., \& Weinberg, M. D. (1984) Dynamical
Friction in Spherical Systems, MNRAS, \textbf{209}, 729 

\bibitem[van der Marel(2004)]{van04} 
van der Marel, R. P. (2004) Intermediate-mass Black
Holes in the Universe: a Review of Formation Theories and
Observational Constraints, in \textit{Coevolution of Black Holes and
Galaxies}, Carnegie Observatories Astrophysics Series, Vol. I.,
ed. L. C. Ho, (Cambridge University Press, Cambridge) p. 37 

\bibitem[Vollmer et al.(2000)]{vol00} 
Vollmer, B., et al. (2000) The Consequences of Ram
Pressure Stripping on the Virgo Cluster Spiral NGC 4522, A\&A,
\textbf{364}, 532 

\bibitem[Vollmer et al.(2001)]{vol01} 
Vollmer, B., et al. (2001) Ram Pressure Stripping and
Galaxy Orbits: The Case of the Virgo Cluster, ApJ, \textbf{561}, 708 

\bibitem[Wang, Cowie, \& Barger(2004)]{wan04} 
Wang, W.-H., Cowie, L. L., \& Barger, A. J. (2004) An
850 Micron SCUBA Survey of the Hubble Deep Field-North GOODS Region,
ApJ, \textbf{613}, 655 

\bibitem[Ward(2003)]{war03} 
Ward, M. (2003) X-ray Components in Spiral and
Star-forming Galaxies, in \textit{Frontiers of X-ray Astronomy},
eds. A. C. Fabian, K. A. Pounds, \& R. D. Blandford, (Cambridge
Univ. Press, Cambridge) p. 117 

\bibitem[Weinberg(1985)]{wei85} 
Weinberg, M. D. (1985) Evolution of Barred Galaxies by
Dynamical Friction, MNRAS, \textbf{213}, 451 

\bibitem[Whitmore \& Schweizer(1995)]{whi95} 
Whitmore, B. C., \& Schweizer, F. (1995) Hubble Space
Telescope Observations of Young Star Clusters in NGC 4038/4039, the
'Antennae' Galaxies, AJ, \textbf{109}, 960 

\bibitem[Whitmore et al.(1999)]{whi99} 
Whitmore, B. C., et al. (1999) The Luminosity Function
of Young Star Clusters in "the Antennae" Galaxies (NGC 4038-4039), AJ,
\textbf{118}, 1551 

\bibitem[Wolter \& Trinchieri(2004)]{wol04} 
Wolter, A., \& Trinchieri, G. (2004) A Thorough Study
of the Intriguing X-ray Emission from the Cartwheel Ring, A\&A,
\textbf{426}, 787 

\bibitem[Xu, Sun, \& He(2004)]{xu04} 
Xu, C. K., Sun, Y. C., \& He, X. T. (2004) The Near
Infrared Luminosity Function of Galaxies in Close Major-Merger Pairs
and the Mass Dependence of the Merger Rate, ApJ, \textbf{603}, L73 

\bibitem[Yan et al.(2004)]{yan04a} 
Yan, L., et al. (2004) High-redshift Extremely Red
Objects in the Hubble Space Telescope Ultra Deep Field Revealed by the
GOODS Infrared Array Camera Observations, ApJ, \textbf{616}, 63 

\bibitem[Yan \& Thompson(2003)]{yan03} 
Yan, L., \& Thompson, D. (2003) Hubble Space Telescope
WFPC2 Morphologies of K-selected Extremely Red Galaxies, ApJ,
\textbf{586}, 765 

\bibitem[Yan, Thompson, \& Soifer(2004)]{yan04b} 
Yan, L., Thompson, D., \& Soifer, T. (2004) Optical
Spectroscopy of K-selected Extremely Red Galaxies, AJ, \textbf{127},
1274 

\bibitem[Young et al.(1996)]{you96} 
Young, J. S., et al. (1996) The Global Rate and
Efficiency of Star Formation in Spiral Galaxies as a Function of
Morphology and Environment, AJ, \textbf{112}, 1903 

\bibitem[Zabludoff et al.(1996)]{zab96} 
Zabludoff, A. I., et al. (1996) The Environments of
"E+A" Galaxies, ApJ, \textbf{466}, 104 

\bibitem[Zabludoff \& Mulchaey(1998)]{zab98} 
Zabludoff, A. I., \& Mulchaey, J. S. (1998)
Hierarchical Evolution in Poor Groups of Galaxies, ApJ, \textbf{498},
L5 

\bibitem[Zaritsky \& Harris(2004)]{zar04} 
Zaritsky, D., \& Harris, J. (2004) Quantifying the
Drivers of Star Formation on Galactic Scales. I. The Small Magellanic
Cloud, ApJ, \textbf{604}, 167 

\bibitem[Zhang(2003)]{zha03} 
Zhang, X. (2003) Secular Evolution of Spiral Galaxies,
JKAS, \textbf{36}, 223 

\bibitem[Zwicky(1959)]{zwi59} 
Zwicky, F. (1959) Multiple Galaxies, Handbuch der Phy,
\textbf{53}, 373 



\end{thebibliography}
